\documentstyle[psfig]{aa}

\input epsf         
\begin{document}

\title{Evolution of the X-ray
spectrum in the flare model \\ 
 of Active Galactic Nuclei}

\author{Suzy Collin\inst{1}, S\'everine Coup\'e\inst{1}, 
Anne-Marie Dumont\inst{1}, Pierre-Olivier Petrucci\inst{2}, 
Agata R\'o\.za\'nska\inst{3}}

\offprints{Suzy Collin (suzy.collin@obspm.fr)}

\institute{$^1$LUTH, Observatoire de Paris, Section de
Meudon, 92195 Meudon, France\\
$^2$Laboratoire d'Astrophysique de Grenoble, 414 rue de la Piscine, 
38041 Grenoble Cedex 9, France\\     
$^3$N. Copernicus Astronomical Centre, Bartycka 18, 00-716 Warsaw, 
Poland}

\date{Received 19 September 2002 / Accepted  19 December 2002 }

\titlerunning{Flare model of accretion discs}
\authorrunning{S. Collin et al.}

\abstract{Nayakshin \& Kazanas (2002) have considered the 
time-dependent 
illumination of an accretion disc 
in Active Galactic Nuclei, in the lamppost model, where it is assumed 
that
 an X-ray source illuminates the whole inner-disc region in 
a relatively steady way. We extend their study to the flare model, 
which 
postulates the release of a large X-ray flux above a small region 
of the accretion disc. A fundamental difference to the lamppost 
model is 
that the region of the disc below the flare is 
not illuminated before the onset of the flare. 
After the onset, the temperature and the ionization state of 
the irradiated skin respond immediately to  
the increase of the continuum, but 
pressure 
equilibrium is achieved later. A few typical test models show that the reflected
 spectrum that follows immediately the increase in continuum flux should always 
 display
 the characteristics of a highly illuminated but dense
gas, i.e. very
intense X-ray emission lines and ionization edges in the soft X-ray 
range. The behaviour of the iron line is however different in
 the case of a ``moderate" and a ``strong'' flare: for a moderate 
 flare, the spectrum displays a  neutral 
component of the
Fe K$\alpha$ 
line at 6.4 keV, gradually leading to more highly ionized  
lines. For a strong flare, the lines are already emitted by
FeXXV (around 6.7 keV) after the onset, and are very 
intense, with an equivalent width of several hundreds eV. A strong flare is 
also characterized by a steep soft X-ray spectrum. The variation timescale 
in the flare model is likely smaller than in the lamppost model, due 
to the smaller dimension of the emission region, so the timescale for pressure equilibrium 
is long compared 
to the duration of a flare. It is therefore highly probable that several flares contribute at the 
 same time
 to the luminosity. We find that the observed 
correlations between $R$, $\Gamma$, and the X-ray flux are well 
accounted for by a combination of flares having not achieved pressure 
equilibrium, also strongly suggesting that the observed spectrum is always dominated 
by regions in non-pressure equilibrium, typical of the onset 
of the flares.
Finally, a flare being confined to a small region of the disc,
the spectral lines should be narrow (except for a weak Compton broadening) and Doppler shifted,
 as stressed 
by Nayakshin \& Kazanas (2001).  All these features 
should constitute
specific variable signatures of the flare model, distinguishing it 
from the 
lamppost model. It is however difficult, on the basis of the present 
observations and models, to conclude in favor of one of the
hypothese.

\keywords{Accretion, accretion discs | galaxies: active |
galaxies: nuclei}}

\maketitle

\section{Introduction}

The model of ``irradiated accretion disc'' is now widely accepted as an
explanation for the presence in the X-ray spectrum of Active Galactic
Nuclei (AGN) of a Fe K$\alpha$ line and an X-ray hump (Pounds et
al. 1990), and for the absence of time delay (or for the very short
delay) between the variations of the optical and UV continuum fluxes
(Collin-Souffrin 1991, Clavel et al. 1992). Both phenomena are most
likely produced by an X-ray flux irradiating a cold medium (the accretion
disc?), which leads to a signature as reprocessed radiation in the UV
range, and to a Fe K$\alpha$ line and a Compton reflection hump in the
X-ray range.

Three different models have been proposed to account for these
properties: an X-ray source (whose origin is unknown) at a relatively
high altitude above the disc, illuminating a large central region (the
``lamppost''), a hot corona completely covering a cold disc (Haardt \&
Maraschi 1991, 1993), and finally a ``patchy corona'' (Haardt et al. 1994)
partially covering the disc.  This last model is often preferred for
various reasons: the shape of the reflected continuum, the correlation
between the Fe K$\alpha$ and the continuum, the absence or weakness of
the helium- and hydrogen-like Fe lines (cf.  Nayakshin 2000, Nayakshin \&
Kallman 2001), the fact that it allows us to account for the large UV over X
observed luminosity ratio (cf.  Walter et al. 1994). It is called the
``magnetic flare'' model, as it refers to magnetic loops reconnecting and
 suddenly releasing their energy as X-ray photons on a small portion of
the disc.

There is presently no clear consensus concerning the physics of the flare,
which is generally considered to be analogous to the sun. In particular, the
height of a flare above the disc, its luminosity and its compactness are
free parameters in addition to others that can be chosen in order to fit the
observed variational and spectral properties of the objects. Poutanen \&
Fabian (1999) and Merloni \& Fabian (2001) have proposed a
model accounting for the spectral variability of accreting black holes
(it is scaled for galactic sources, but it applies to AGN as well).  
Many simultaneous flares are required to account for the observed
luminosity of an AGN. To avoid the smearing of strong luminosity variations,
 they propose that the amplitude of variability can
be explained if these flares are triggered in a chain reaction. Each
flare produces an ``avalanche'' of neighbouring flares, whose
duration and amplitude determine the timing properties. In
particular, the distribution of size and luminosity of the avalanches accounts
for the shape of the power spectrum density, which gives the power
 as a function of the Fourier frequency  (PSD, see also Zycki
2002). The most powerful avalanches have the largest size (corresponding to the
largest time scales), and they correspond to the ``break'' and the red branch
of the PSD (cf.  below).  In this model, the height of a magnetic loop
increases during the avalanche, increasing the number of UV photons
available for the cooling of the corona and causing the observed
softening of the X-ray spectrum. It should be at least one order of
magnitude larger than the pressure scale height of the disc. Consequently
an avalanche can illuminate a relatively large fraction of the inner disc
(since the radius of the illuminated region is of the order of the height
of the source). 
 Nayakshin \& Kazanas (2001) argue on the contrary that
the height of a flare should be of the order of the scale height of the
 disc, and that an avalanche of flares should 
  illuminate only a very small fraction of the
disc. In both cases the total number of simultaneous powerful 
avalanches is small, so that the total area illuminated by them is small
compared to the disc area.  This is a fundamental issue as {\it it implies
that the region of the disc located under an avalanche was not illuminated
before the onset of the avalanche}.

 Our aim in this paper is to study the
implication of this model, in particular on the soft X-ray spectrum and
on the Fe lines. Note that in the following we will use the word 
``flare", even if it is actually an avalanche.

Several recent results concerning time variations of the Fe 
lines are very intriguing in the context of irradiated discs. The
expectation would be that the Fe lines respond to variations of
the X-ray continuum flux in a simple way, in the same way as the broad optical-UV lines
respond to the UV-X continuum flux. This is far from being the case, and
the behaviour of the lines is actually very complex. The interpretations
of the results are sometimes even contradictory, for instance for
MCG -6-30-15 (Lee et al. 1999, Vaughan \& Edelson 2001), where it is not clear
 whether the line flux stays
constant, while the continuum varies.  Another example is Mkn 841, where
a rapid  (less than 15 hours) and strong (a factor two) variation of the
 Fe K$\alpha$ line is related only to a weak variation (less than 20 $\%$)
 of the X-ray continuum (Petrucci et al. 2002).

Moreover the variations of the optical/UV continuum, which are generally
assumed to be induced by variations of the X-ray flux and should
therefore follow them after a short time delay if they are emitted by the
accretion disc (Rokaki et al. 1993, Kazanas \& Nayakshin 2001), do not
obey a simple relationship with the X-ray light curves. For instance in
NGC 7469, the UV flux leads the 2-10 keV flux in the peaks of luminosity
(Nandra et al. 1998) while the 2-10 keV flux leads the UV flux during the
minima of luminosity. However this effect can be an artefact due to the
limited range of energy considered: since the X-ray spectrum becomes
softer when the UV flux is large, the X-ray flux might be reprocessed
below 2 keV. In NGC 3516 there is no clear correlation between the UV and
the X-ray flux (Edelson et al. 2000).

It is clear that the response of the irradiated medium is not simply
the consequence of the variation of the ionization
  state  and of the thickness of the ``hot skin''
 of the disc irradiated by a variable X-ray flux, as discussed
either for a constant density medium or for a 
medium in 
hydrostatic equilibrium (Zycki \& 
R\'o\.za\'nska 2001). 
A most interesting idea stressed for the first time by Nayakshin \&
Kazanas (2002, hereafter NK02) is that if the illuminating flux
varies on short time scales, the Fe line flux should not be a
function of the instantaneous illuminating spectrum. In response to a
variation of the illuminating flux, the atmosphere of the accretion disc
remains  during a relatively long time, required for readjustment of
the hydrostatic balance, in a non-equilibrium state.

\begin{figure}
\begin{center}
\psfig{figure=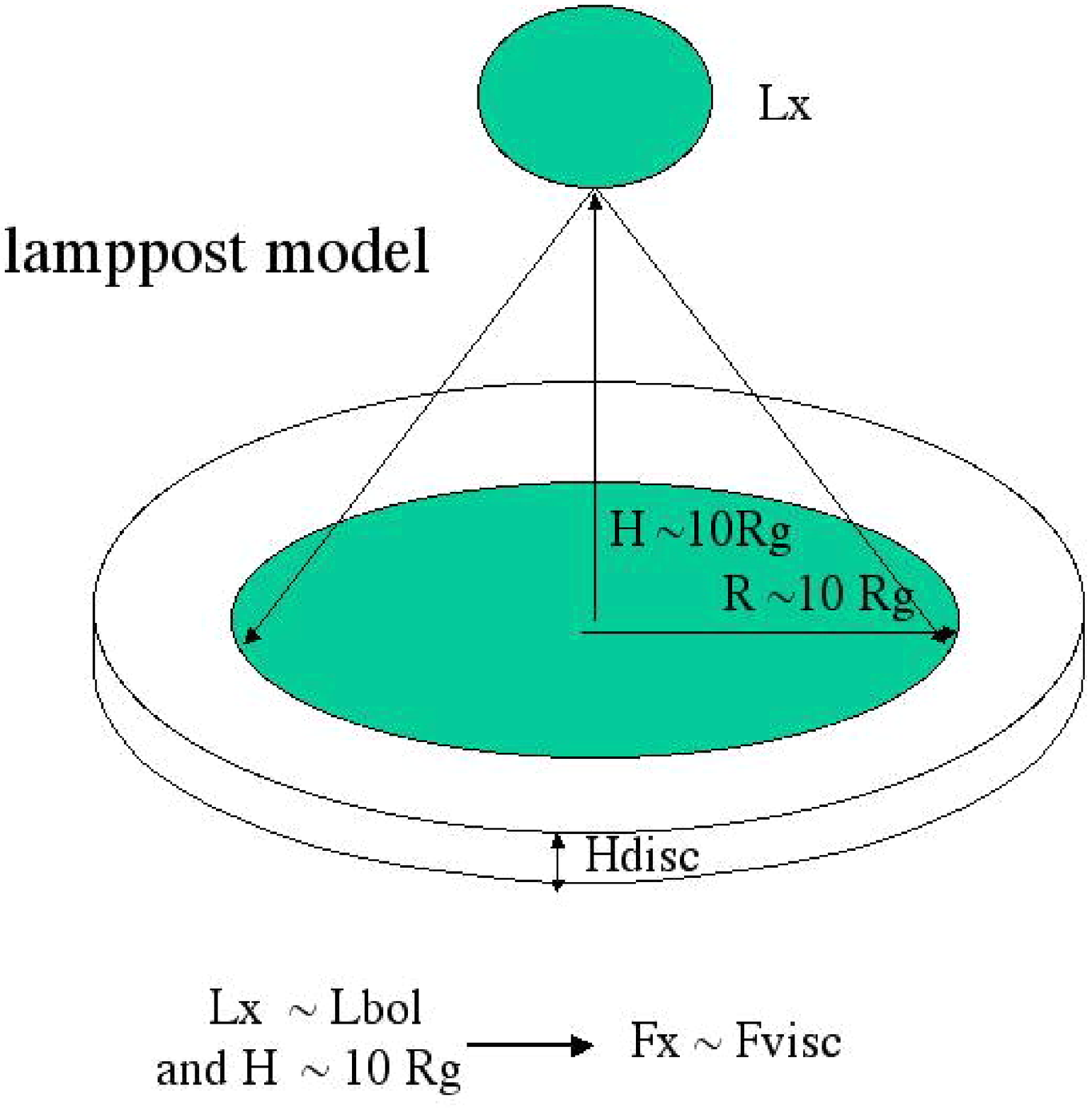,width=9cm}
\psfig{figure=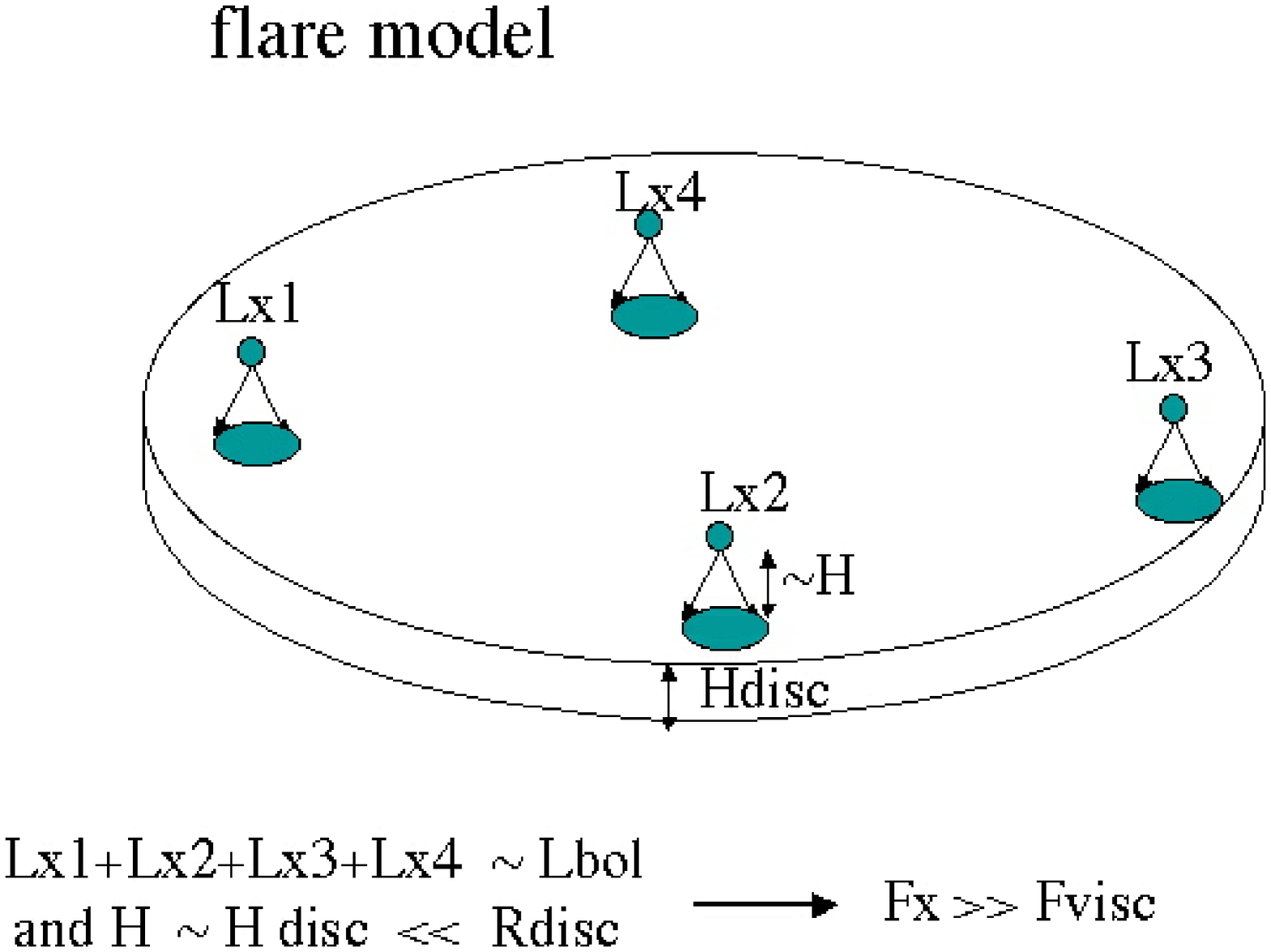,width=9cm}
\caption{ The lamppost and the flare models. }
\label{fig-lamppost-flare}
\end{center}
\end{figure}

In this paper we extend this idea, discussing the fate of a 
region of the disc illuminated by a flare, a case not 
considered 
by NK02, who limited their study to the ``lamppost model''. To 
clarify the discussion, Fig. \ref {fig-lamppost-flare} shows our 
vision of the lamppost and the flare models.  

In both cases, the disc is 
illuminated by an X-ray source whose luminosity is equal to a 
fraction of the bolometric luminosity, located at a given height $H$ 
above the disc. The 
lamppost is located in the center (above the black hole), and its height 
is of 
the order of 10-20 $R_{G}$ ($R_{G}$ being the gravitational radius 
$GM/c^2$), while a flare is located at any radius, and its height 
above the disc is much smaller. Several flares can be shining 
at the same time. The dimension of the region
 illuminated by a source located above the 
disc is of the order of its height, and the flux on this 
region is of 
the order of $L/4\pi H^{2}$. For a flare, the flux is thus much larger 
 than for a lamppost. It is also much larger than the 
viscous flux provided by the release of gravitational energy, which 
is maximum at a few $R_{G}$ and
decreases as $R^{-3}$ (where $R$ is the distance to the center).

 According to 
this 
discussion, we assume therefore that the 
flare model differs from 
the lamppost model in three respects:

\begin{itemize}

\item In the lamppost model, the disc is illuminated permanently. The
illuminating flux can switch from a state $F_{X1}$ to a state $F_{X2}$,
with $F_{X2}\sim$ a few $F_{X1}$, or inversely, but the illumination is
{\it always} present.  Therefore an irradiated layer, called the ``hot
skin'', always exists above the inner disc. However the hydrostatic
equilibrium corresponding to the two states is different, and the hot
skin as well as its spectrum will evolve slowly (cf. later) 
from one equilibrium state to the
other.  On the contrary, in the flare model, there is {\it no
illumination} in the initial state (and in the final state, after the
extinction of the flare), so the newly formed hot skin is much denser
than an illuminated atmosphere (it does not exclude that another part of
the disc is already illuminated, as we will discuss in Section 4).

\medskip
\item In the lamppost model, the disc is illuminated by a flux $F_X$
smaller than or of the order of the viscous flux $F_d$, while in the 
flare model,
 $F_X$ is much larger than $F_d$.
 
 \medskip
 \item In the lamppost model, the whole inner disc is illuminated, 
 while in the flare model the illumination acts only on a small fraction
of the disc. 
 
\end{itemize}

In the following section, we recall the relevant time scales, and in 
Section 3 we give results corresponding to a few typical cases. 
Implications for the 
observations are discussed in Section 
4.

\section{Physical parameters}

\subsection{Time scales}

When the X-ray flux varies, the illuminated medium responds with 
different characteristic times. Though they are recalled by 
NK02, we give them here in the context of the 
present study.

\medskip
\noindent 1. The radiation transfer time, $t_{rt}$. In a completely
ionized medium, $t_{rt}= \tau_{es}(1+\tau_{es})/(\sigma_{T}cn)$,
where $\tau_{es}$ is the Thomson thickness of the reflecting medium,
$\sigma_{T}$ is the Thomson cross section, and $n$ is the electron
density (note that resonance line scattering can also play a role, cf.
below).  Since X-ray photons are absorbed in a few Thomson units, the
optical thickness of the hot skin is smaller than or of the order of
unity, and therefore $t_{rt}\le 100 n_{12}^{-1}$ s, where $n_{12}$ is
the density expressed in 10$^{12}$ cm$^{-3}$. We will see that the
average density at the onset of the flare is probably larger than
10$^{12}$ cm$^{-3}$. So $t_{rt}$ is small, and the whole atmosphere receives
the new flux quasi-simultaneously.

\medskip
\noindent 2. The time for readjustment of the ionization equilibrium, 
which is 
the longer of the ionization time, $t_{ion}$, and the recombination 
time, $t_{rec}$. One has
 $t_{ion}\sim <h\nu >/(F_X\sigma_{ion})$, where $<h\nu >$ is the 
mean energy of the ionizing photons, and $\sigma_{ion}$ is the 
ionization 
cross-section. 
For the physical conditions holding near the surface of the accretion disc,  
$t_{ion}\sim 10^{-7}F_{16}^{-1}$ s, 
where the flux is expressed in 10$^{16}$ ergs cm$^{-2}$ s$^{-1}$, a 
typical 
value for $F_X$ during a flare 
(cf. Nayakshin et al. 2000, Ballantyne \& Fabian 2001, and below).
  $t_{rec}\sim 1/(n\alpha_{rec}) \sim 10 n_{12}^{-1}$ s, where 
$\alpha_{rec}$ is the recombination coefficient of the ion under 
consideration. So 
again these times are short, and one can consider that the medium 
readjusts
instantaneously to the new conditions.

\medskip
\noindent 3. The time for thermal equilibrium, $t_{therm}$, i.e. the time
required to radiate the energy of the medium, $nkT/(n^2 \Lambda)$, where
$\Lambda$ is the cooling function. At the onset of a flare, $\Lambda$ is
dominated by atomic processes and is of the order of 10$^{-23}$ erg
cm$^{+3}$ s$^{-1}$. This gives $t_{therm}\sim 10T_6
n_{12}^{-1}\Lambda_{23}^{-1}$ s, again a small value \footnote{If the
energy balance is dominated by optically thick resonance lines, one
should take into account the diffusion time for resonant scattering,
which is long.  It is indeed the case of the deep layers for the
atmosphere, but the temperature stays almost constant in these layers}.

\medskip
\noindent 4. The time for readjustment of the hydrostatic equilibrium
(the dynamical time).  Actually this time is of the order of that
necessary to create the hot skin, $t_{dyn}=H/c_{s}$, where $H$ is the
scale height of the atmosphere (after hydrostatic equilibrium), and $c_s$
is the sound velocity in the hot skin (after thermal equilibrium). Indeed the hydrostatic
equilibrium of the disc itself is almost not changed by the illumination
 if $F_X/F_d \ll \tau_{disc}$ (except possibly close to the surface), where
$\tau_{disc}$ is the optical thickness between the surface and the equatorial
plane (cf. Hubeny 1990, Hur\'e et al. 1994), a condition easily fulfilled
as the disc is optically very thick for typical accretion rates  in 
quasars and Seyfert galaxies. For a Thomson thickness of the
atmosphere of order of unity, $t_{dyn}\sim 10^5 T_6^{-1/2} n_{12}^{-1}$
s, if one assumes that the readjustement takes place at the sound
speed. It is possible that shock waves with a
relatively high Mach number contribute to this readjustment, so 
the dynamical time could be smaller. Even in this
case, $t_{dyn}$ would be much larger than the previous ``microscopic"
time scales.
 
 \begin{figure}
\begin{center}
\psfig{figure=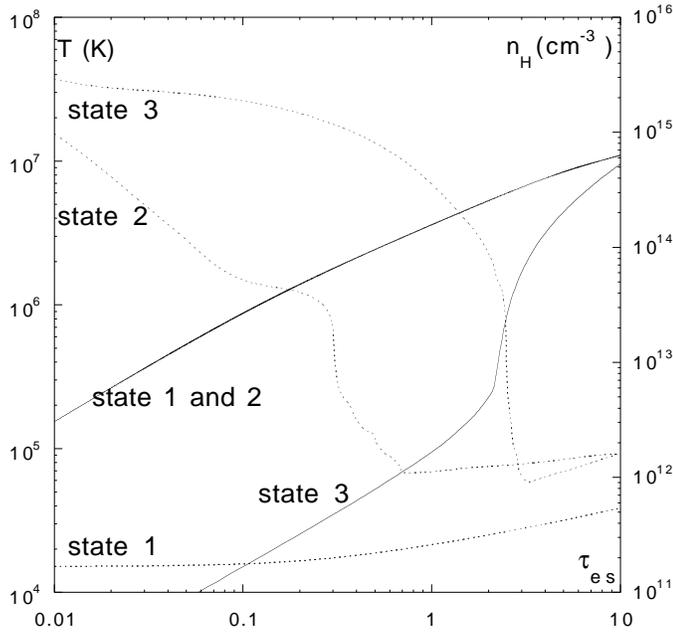,width=9cm}
\caption{Temperature and density versus $\tau_{es}$ for model H1, 
before (state 1), after the onset of 
 the flare
(state 2), and 
after pressure equilibrium is reached (state 3);
 dotted line: temperature; solid
 line: density. We recall that state 2 is reached at a time
$t_{grow}\sim 10^3$ s after the beginning of the flare, while state 3 is achieved 
after a much longer time, $t_{dyn}\sim 10^5$ s.}
\label{fig-hydro-T-n}
\end{center}
\end{figure}

The onset of a flare is not instantaneous, and all these 
times should
 be compared
 to the
 variability time scale of the 
illuminating continuum. Three time scales are of interest in this 
context. 

First, one has to take into account the ``growing" time  of a flare
(or equivalently the ``fading" time) 
as seen 
from the underlying disc, $t_{grow}$. It is probably of the order of 
the vertical extension of an individual flare divided
by the Alfven velocity. One can assume 
that the 
dimension of a magnetic loop is of the order of the scale height of 
the disc
(Nayakshin \& Kazanas 2001). In the standard model of accretion 
discs 
(Shakura \& Sunayev 1973) where the inner regions are dominated by radiation 
pressure 
and Thomson opacity, this scale height is roughly equal 
to 10$^{13} 
M_8 \dot{m}_{-1}$ cm, where $\dot{m}_{-1}$ is the bolometric to 
Eddington luminosity
 ratio in units 0.1, 
and $M_8$ is the black hole mass in 
10$^8$M$_{\odot}$. Thus  $t_{grow}$ should be of the order of
10$^3$ s. 

Second, one should consider the time scale corresponding to the 
{\it observed} increase of the 
X-ray continuum, 
$t_{fl}$, i.e. to the onset of the whole avanlanche. It
 is difficult to predict theoretically, but one knows that
 the PSD extends from 10$^{-5}$ to 10$^{-2}$ 
Hz, and consists of two power laws and a break at $\sim 3\ 10^{-4}$ Hz, 
which  should correspond to the largest and most powerful active 
regions. Accordingly, $t_{fl}$ is
 larger than, or of the order of $t_{grow}$, and smaller than the 
dynamical time. One deduces also that the radius $R_{fl}$ of
powerful flares is smaller than 
10$^{14}$ cm. It is therefore smaller than the radius of 
the inner accretion
 disc giving the bulk of the viscous dissipation, $\sim 20 R_G = 3\ 
10^{14}M_8$ cm. 
Note that it is in agreement with the conclusion of Nayakshin \& 
Kazanas 
(2001)
based on a different argument. 

The third timescale is the whole duration of a flare, for which we 
have no real indication, and which could well be as large as 10$^{5}$ s 
(the largest 
time given by the PDS).

\medskip

To summarize, one has: 

\begin{equation}
t_{microscopic} \ll t_{grow}\sim 10^3\ {\rm s} \le t_{fl}\ll 
t_{dyn} 
\sim 10^5\ {\rm s}.
\label{eq-comp-temps}
\end{equation}

Compared with the sampling time of an observation, the onset 
of the flare
 can thus be observed.  The temperature and the ionization equilibrium
 respond immediately to an 
 increase of flux, but the density structure stays unchanged. The 
adjustement in
density (i.e. in pressure) 
is much longer, and might not be achieved before the flare ends, or 
before another flare has begun. So it is likely that the observed 
spectrum 
corresponds to a mixture of spectra in different non-equilibrium
phases.

An important point to notice is the difference in timescales 
between the lamppost and the flare models, simply due to the larger 
size of the illuminated region in the lamppost model: while a typical 
time for the onset of a flare should be $10^{3}$ s, it should be 
at least one order of magnitude larger for the lamppost model.  We 
recall that all these times refer to a black hole mass of 10$^8$M$_{\odot}$.

\subsection{Fluxes}

\begin{figure*}
\begin{center}
\psfig{figure=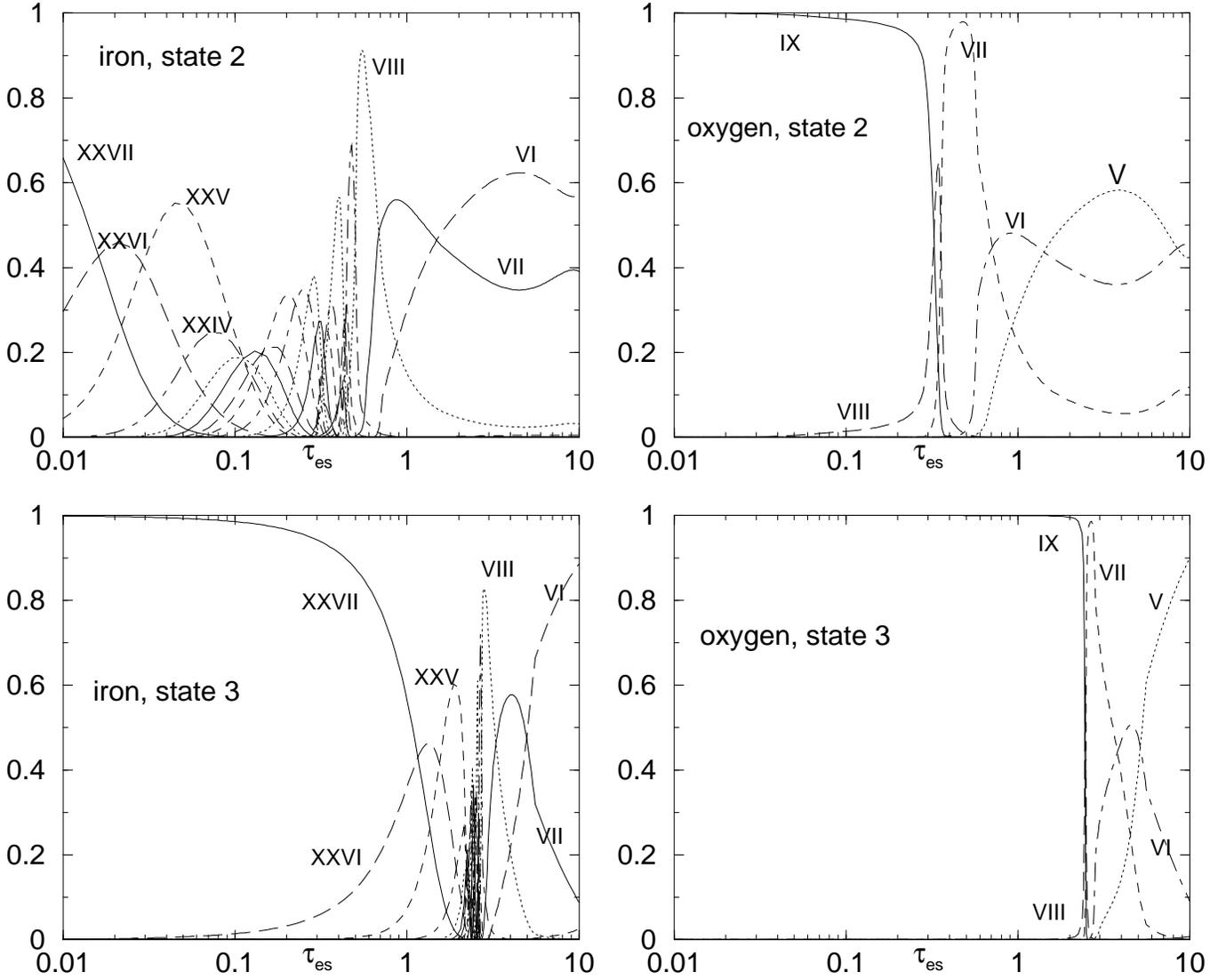,width=18cm,angle=270}
\caption{Fractional ionization of oxygen and iron for model H1.}
\label{fig-hydro-deg-O-Fe}
\end{center}
\end{figure*}

The flare  model is characterized by the fact that most of the disc receives a very
 small illumination, 
whereas 
one  region is suddenly illuminated by an X-ray flux $F_X$
 much larger 
than the underlying viscous flux $F_d$. So one can neglect the incident flux 
 before the flare.

To account for the observed X-ray variability, a flare
 should have a luminosity comparable to the 
average 
X-ray luminosity,
 which is itself a fraction  $f$ of the bolometric luminosity
(typically $f\sim 10\%$). It  means 
that the illuminating flux in the region covered by the flare 
should be of the order of: 
\begin{equation}
F_{X}\sim  10^{17} 
f_{-1}\dot{m}_{-1}M_8^{-1}r_{fl}^{-2}\ \ {\rm erg\ cm}^{-2}{\rm 
s}^{-1}
\label{eq-FX-flare}
\end{equation}
 where 
$f_{-1}$ is expressed in units of 0.1, and
 $r_{fl}$ is the size of the flaring region 
expressed in $R_G$. Since the model implies that this size
 is small 
compared to 
the radius of the disc corresponding to the bulk of the ``viscous" 
luminosity, 
say 20$R_G$, the illuminating flux is larger than the 
viscous flux
\begin{equation}
F_d\sim 10^{15} 
\eta_{-1}^{-1}\dot{m}_{-1}M_8^{-1}\left({R_{ill}\over 20R_{G}}\right)^{-3}
\ \ {\rm erg\ cm}^{-2}{\rm s}^{-1}
\label{eq-Fvisqueux}
\end{equation}
 where $ \eta_{-1}$ is the 
efficiency of mass-energy conversion in units 0.1 and $R_{ill}$ is the 
radius of the illuminated region. 
Thus after the onset of the 
flare, one can neglect the viscous flux.

\section{Test models}

It is out of the scope of this paper to solve a real time-dependent 
problem, which would be very difficult to handle, as it would imply a 
dynamical study accounting for
shocks and winds. We only want to show what kind of spectral 
features
 should 
characterize the flare model, when an important variation of the 
X-ray flux is 
detected  (say, a few tens of $\%$).
We will assume that at a time t=0, the disc is
 irradiated by a flux larger than the viscous one, which then 
remains
constant. The thermal and ionization equilibria respond to the 
variation 
 of the illuminating flux only in a few tens seconds.
Then the atmosphere begins to expand 
slowly under the effect of the radiation plus gas pressure. If the 
flare lasts for more than, say, one day, hydrostatic
equilibrium can be achieved. We will simply show the difference of 
spectra 
 a short time after the flare, and after hydrostatic 
equilibrium
 is reached.  To this aim we use several test-models. 

\subsection{Standard disc}

The first 
model
is a standard irradiated $\alpha$-disc in hydrostatic equilibrium in 
the gravitational field of the central black hole. 
The method for solving the transfer and the 
vertical structure of the disc is described in R\'o\.za\'nska et al. 
(2002). 
Briefly summarized, two computations are coupled.
In the inner layers, the vertical structure is 
computed according to R\'o\.za\'nska
et al. (1999); the radiative transfer is treated in the diffusion
approximation with convective transport included; 
the viscous energy dissipation is given by the local 
$\alpha$-viscosity description, with
the viscous energy generation proportional to the total 
pressure. In the superficial 
layers ($\tau_{es} \le 10$),
 the transfer is solved with the photoionization-transfer 
code Titan described in Dumont et al. (2000) and updated by  Coup\'e 
et al.
(2003) to include improved atomic data and a larger 
number 
of transitions. Comptonization is taken into account above 1 keV through the 
coupling with a Monte Carlo code, NOAR, also described in Dumont et al. 
(2000). 
The vertical temperature and the density profiles 
are determined 
 according to hydrostatic equilibrium, both in the 
 disc and in 
the hot skin. When multiple solutions arise in the 
atmosphere,
 the ``hot solution" is chosen (actually this choice has no influence 
on the 
 spectrum for the high flux considered here, cf. Coup\'e et al. 2002).

The test model is an accretion disc around a Schwarzschild black 
hole
of mass $M=10^8$ 
M$_{\odot}$, with an accretion rate $\dot{m}=L/L_{Edd}=0.001$ and a 
viscosity parameter  $\alpha=0.1$ (model H1). Before  the time $t=0$, 
the disc is not illuminated. After the time $t=0$,
the disc is illuminated by a flux equal to 
10$^{15}\ {\rm erg\ cm}^{-2} {\rm s}^{-1}$. 
We assume that the flare takes place at a radius $R$ equal to
18$R_G$, 
corresponding to a viscous flux of 7 $10^{12}\ {\rm erg\ cm}^{-2}{\rm 
s}^{-1}$. 
The spectral distribution of the incident 
continuum is a 
power-law with an
energy index $0.9$, extending 
from 1 eV up to 100 keV. This model will be discussed in detail by 
R\'o\.za\'nska et al.
 in a paper in preparation, mainly aimed at comparing low and high 
illumination of the disc, and their results with those obtained by 
other groups.
Note that the model can correspond to a different set of values of 
$M$, $\dot{m}$, and $R$, as the main parameters which matter in this 
problem are $F_{X}$, $F_{X}/F_{d}$, here equal to 140, and the gravity parameter
$A$ defined 
by Nayakshin et al. (2000), here equal to  0.023 just after the 
flare, and to 0.34 when hydrostatic equilibrium is achieved. 

Fig. \ref{fig-hydro-T-n} displays the vertical distribution of 
density and of 
temperature in the atmosphere, before  the onset of the flare (state 1)
 and after a time 
$t_{grow}$ (state 2).
In state 1, the disc atmosphere is cold  
and relatively dense 
($T\sim 10^4$ K and $n\sim 10^{14}$  cm$^{-3}$ at $\tau_{es}=1$). 
In state 2, the density profile is not changed, but thermal and ionization 
equilibria
are achieved and $T$ reaches 10$^7$ K in a skin of optical thickness
$\tau_{es}\sim 
0.3$. After the much longer time $t_{{dyn}}$, hydrostatic equilibrium is 
achieved (state 3),  the density averaged over the hot skin
 is  smaller ($n\sim 10^{12}$  cm$^{-3}$), and the hot skin 
extends deeper into the disc. It is interesting to note that the temperature profile
 is more abrupt 
when pressure equilibrium is achieved, leading in particular to the 
disappearance of intermediate ionization states. This is well known and
due 
to the thermal instability generally present in a slab having an imposed pressure
 (Ko \& Kallman 1994). 

Fig. \ref{fig-hydro-deg-O-Fe} shows the comparison between the 
fractional ionization of oxygen and iron in states 2 and 3. 
One sees clearly that in state 2 the layer contributing to the bulk 
of the reflection spectrum ($\tau_{es}\sim 0.1$ to 1) is dominated by 
intermediate species of iron and by OVII and OVIII ions,
 while in state 3
it is  dominated by fully ionized iron and oxygen.

\begin{figure}
\begin{center}
\psfig{figure=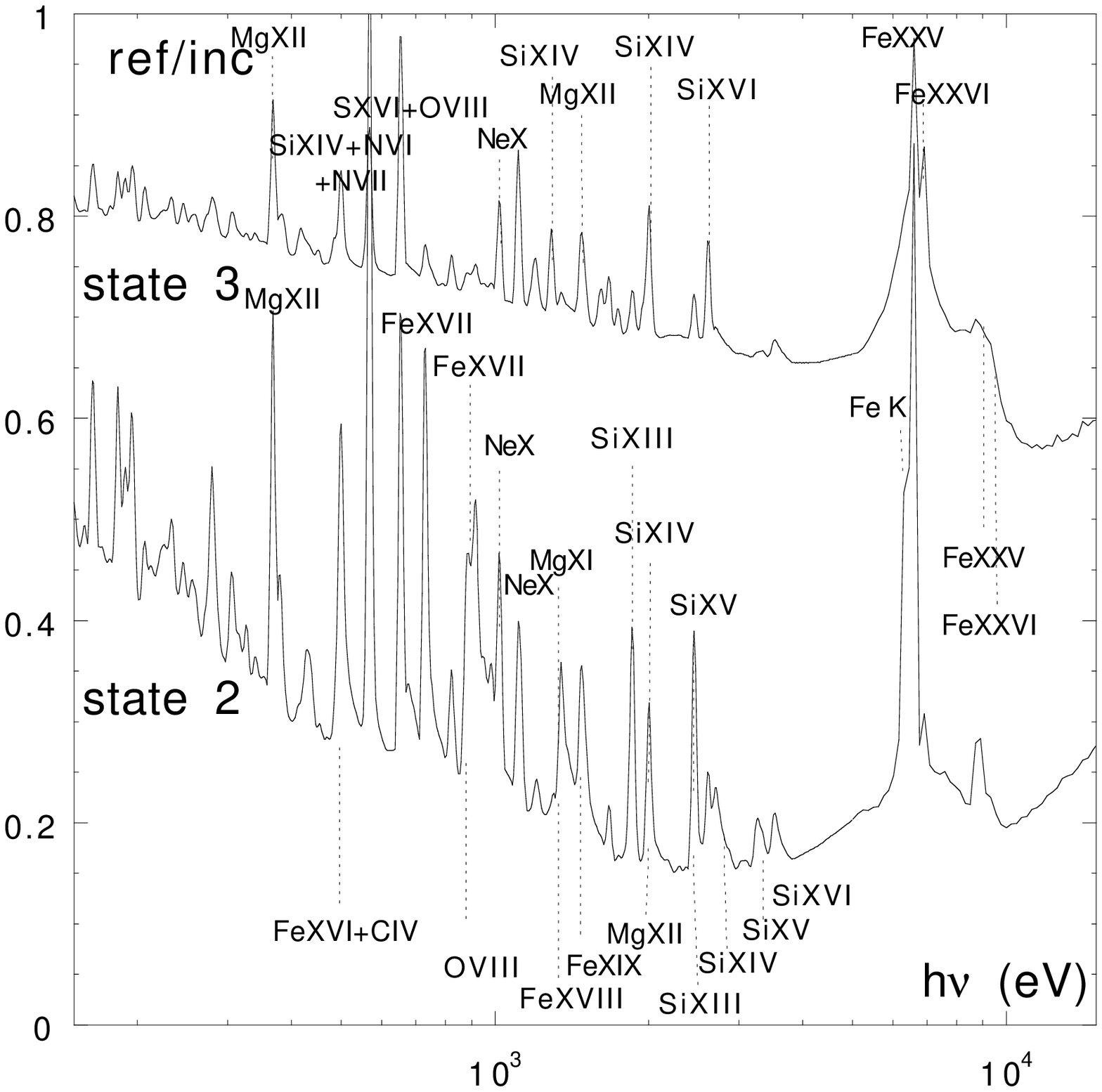,width=9cm}
\psfig{figure=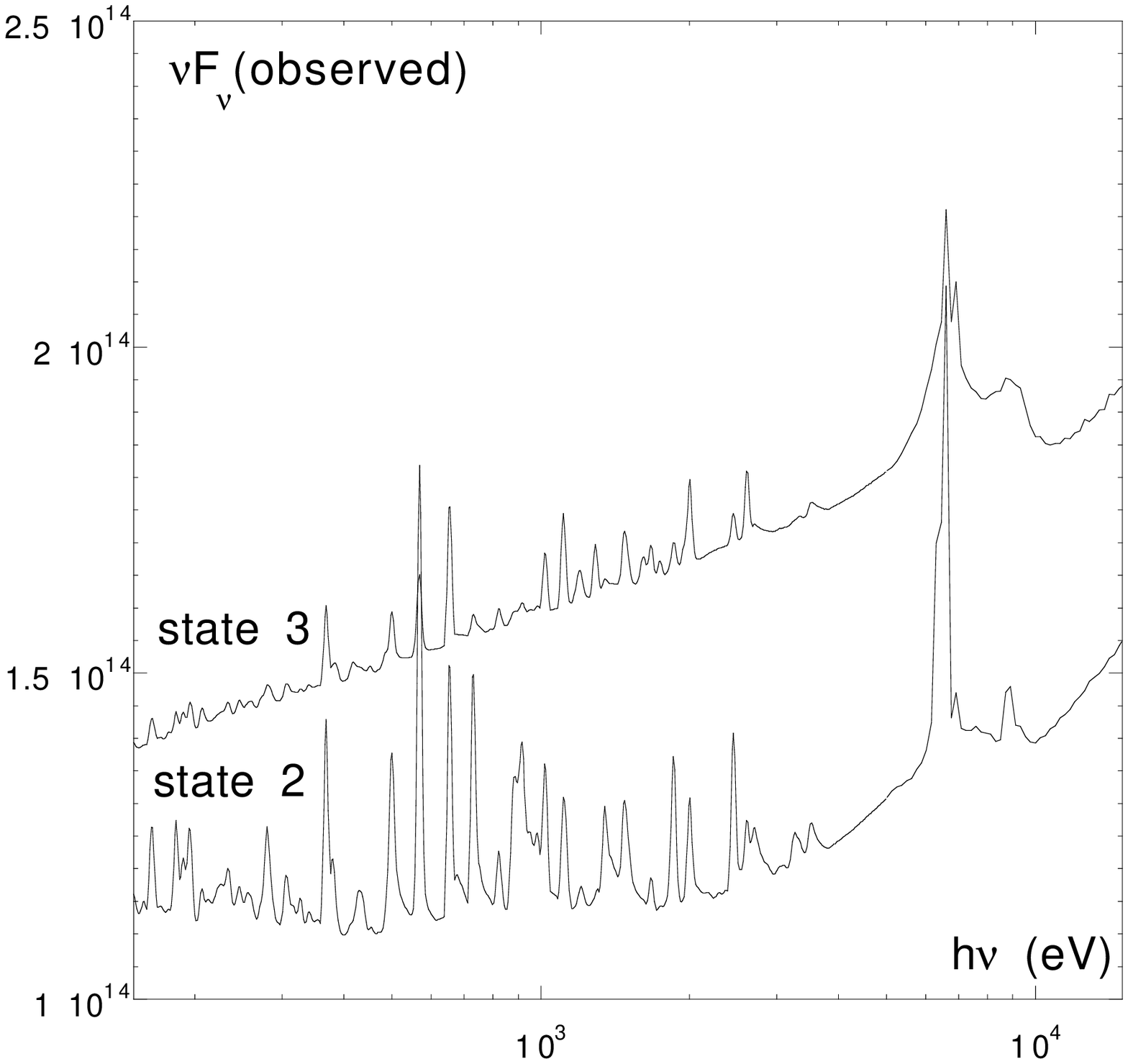,width=9cm}
\psfig{figure=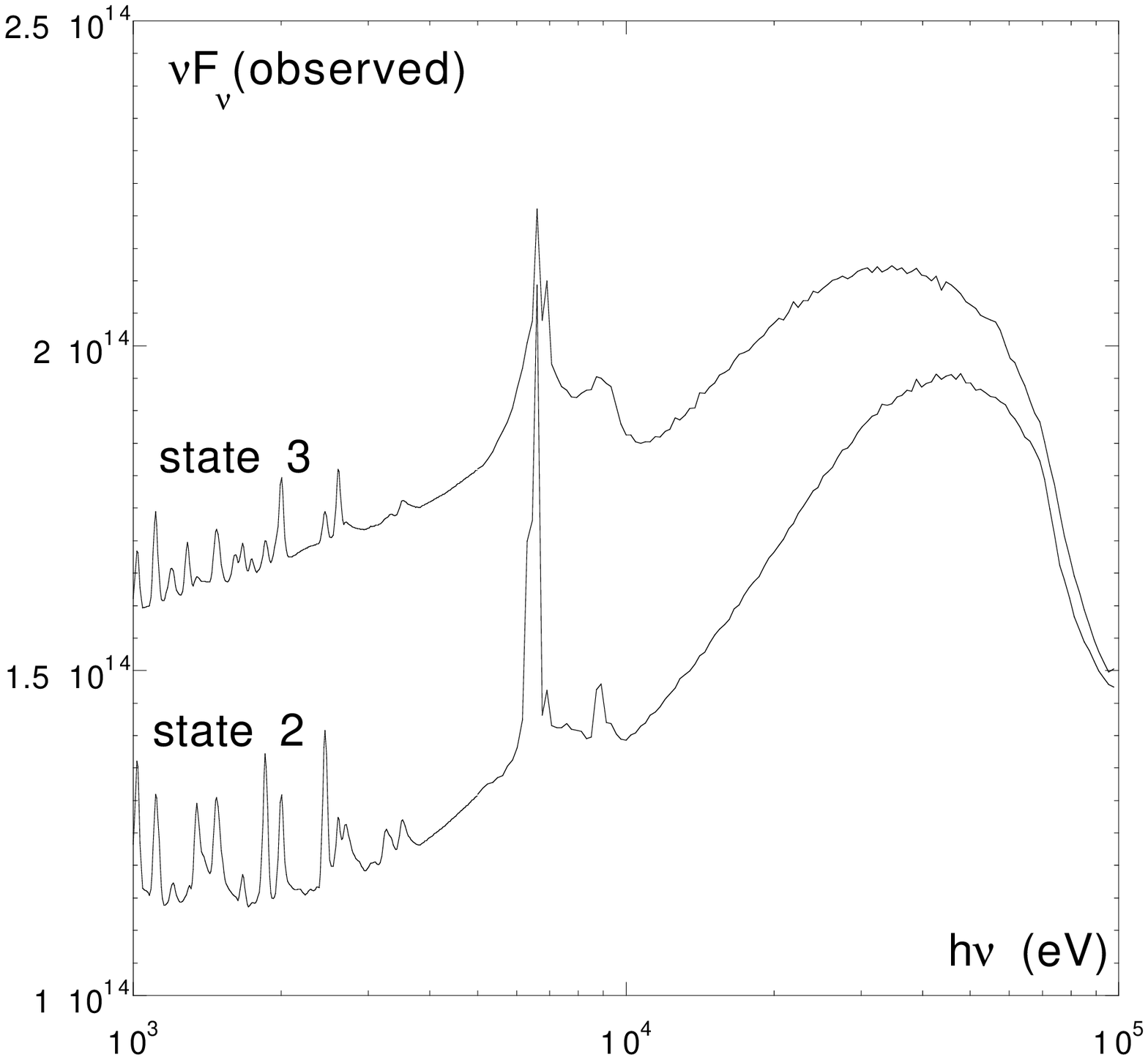,width=9cm}
\caption{Computed spectrum for states 2 and 3 of model H1. 
The top figure gives the ratio of the reflected to the incident 
spectrum, and the two bottom figures give the sum of the two components 
(i.e. the observed spectrum) in two spectral ranges. Some lines and 
edges 
are identified (lines: above the spectrum; edges: below the 
spectrum). The spectra are displayed with a spectral resolution of 44. }
\label{fig-hydro-spe}
\end{center}
\end{figure}

This explains the behaviour of the reflected and observed spectra for states
 2 and 3,  
 shown in 
Fig. \ref {fig-hydro-spe} (we recall that in state 1 there is no reflection 
at all). The top figure  gives the ratio of the reflected to 
the incident 
spectrum, and the two bottom figures show the 
reflected plus incident spectrum in
two spectral bands, to 
perform a direct comparison with the observations \footnote{In the following, 
we will call 
the sum of the reflected plus incident spectrum 
the ``observed spectrum", since it corresponds to the spectrum 
observed for a coverage 
factor of the source equal to 1/2, as it is the case for an irradiated disc.}.

Let us compare the spectra of the two states in the different spectral 
bands:

\medskip
\noindent - In the keV range

In state 2 the gas is not 
highly ionized, so the spectrum displays a Fe K$\alpha$ 
line 
at 6.4 keV, which appears as an intense shoulder in the line profile, 
itself peaking at 6.7 keV, and dominated by the FeXXV line. 
H-like iron is very weak, as is the 
ionization 
edge at 10 keV. 
In state 3, the reflecting layer is thicker, 
the reflected continuum is more intense (in other words the albedo is 
larger), the 6.4 keV line has disappeared, and the FeXXVI Ly$\alpha$ line at
 6.9 keV has appeared, as well as an
 intense ionization edge in absorption at 10 keV. Note also that the 
 iron line displays intense Comptonized wings, in particular on the red 
 side, which are completely absent in state 2.
 
\medskip
\noindent - In the soft X-ray range

In state 2, the spectrum
displays very intense lines and ionization edges in emission, which 
are much weaker in state 3. In this last state, the soft 
X-ray lines should be partly smeared out by Comptonization, which is 
not taken into account in this energy range with our code, so the lines should be 
even less visible \footnote{Below 1 keV Comptonization is taken into account 
in the 
line intensities, but not in the line profiles.}. 

\medskip

It is interesting to note that {\it the continuum is curved and does 
not look like a power-law, even in a small energy range}. In particular 
its shape in the 2-10 keV range varies from state 
2 to state 3, and could well mimic the variation of an extended 
relativistic red wing.

\medskip

The presence of the neutral iron line after the 
onset 
of the flare is caused by the specific model chosen here, where the 
illuminating flux is not very high. In the case of a larger flux,
 the illuminated skin could be highly ionized immediately 
after the 
onset 
of the flare, as we will see in the next section. If the illuminating flux
 reaches a very high local value, the hot 
skin can become
 completely ionized 
and the lines and edges can decrease and even disappear completely
 (cf. Nayakshin \& Kallman 2001).

\subsection{Medium in pressure equilibrium}

\begin{figure}
\begin{center}
\psfig{figure=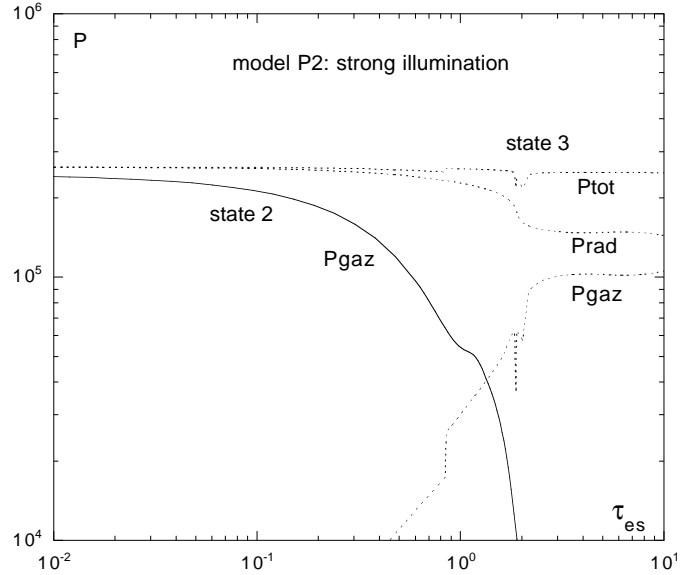,width=9cm}
\caption{Gas and radiative pressure for model
P2; solid line: gas pressure after the onset of 
 the flare, and before pressure equilibrium is reached
(state 2);
 dotted line:  gas, radiative, and total pressure 
after pressure equilibrium is reached (state 3): the total pressure 
is constant and equal to the gas pressure at the surface in state 2.}
\label{fig-P2-P}
\end{center}
\end{figure}

\begin{figure}
\begin{center}
\psfig{figure=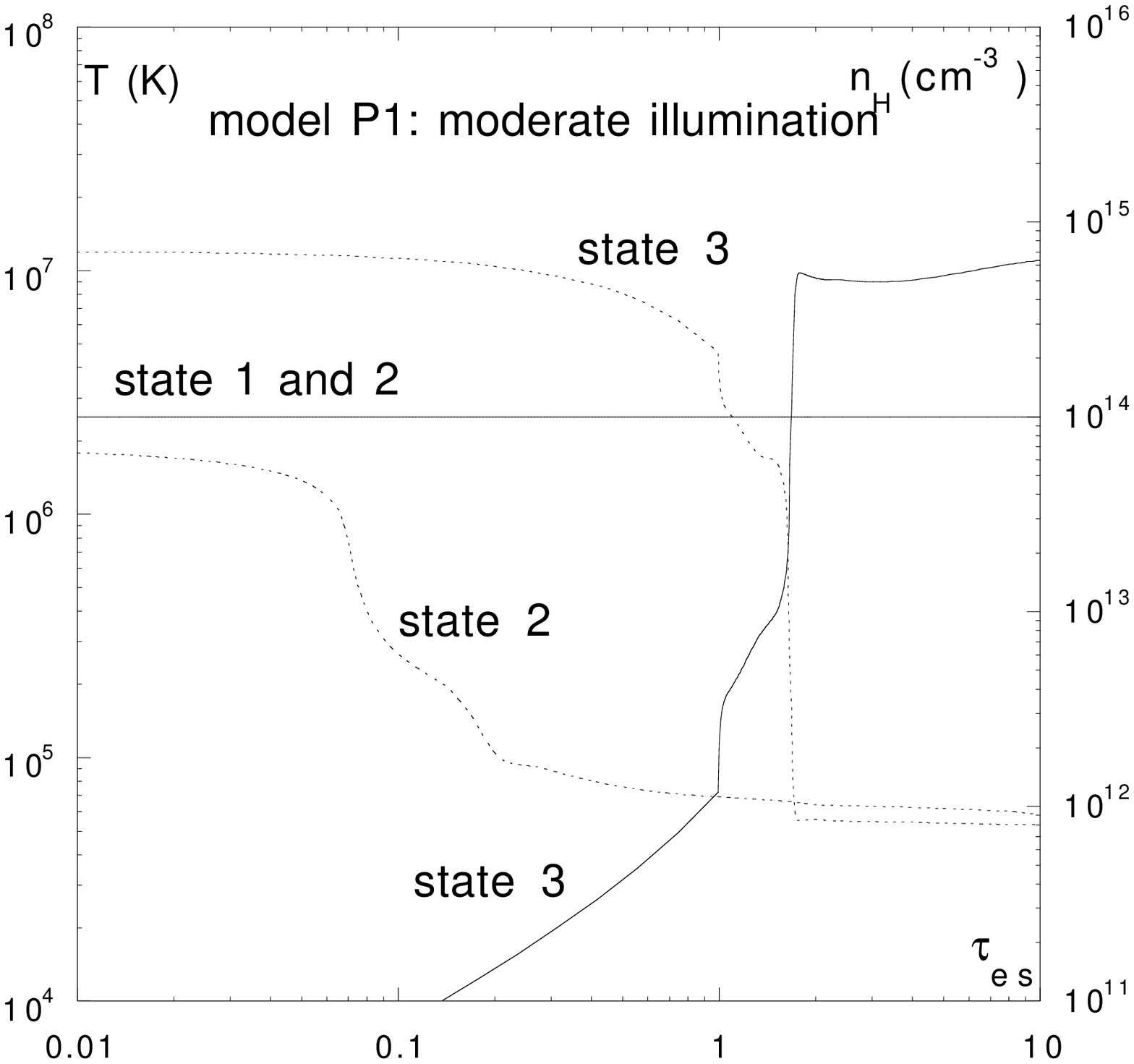,width=9cm}
\psfig{figure=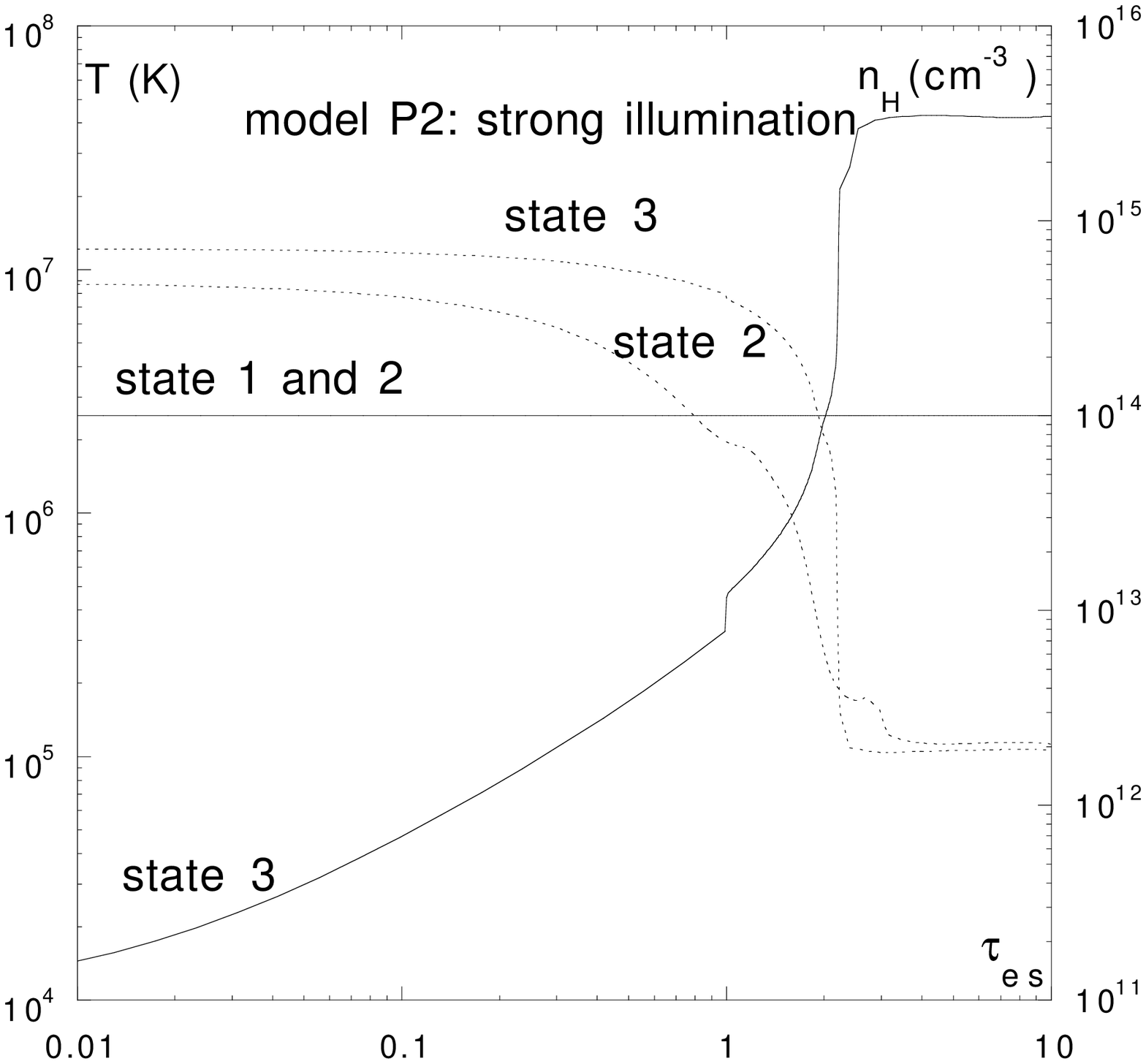,width=9cm}
\caption{Temperature and density versus $\tau_{es}$ for models P1 and 
P2;
 dotted line: temperature; solid
 line: density.}
\label{fig-pcst-T-n}
\end{center}
\end{figure}

\begin{figure}
\begin{center}
\psfig{figure=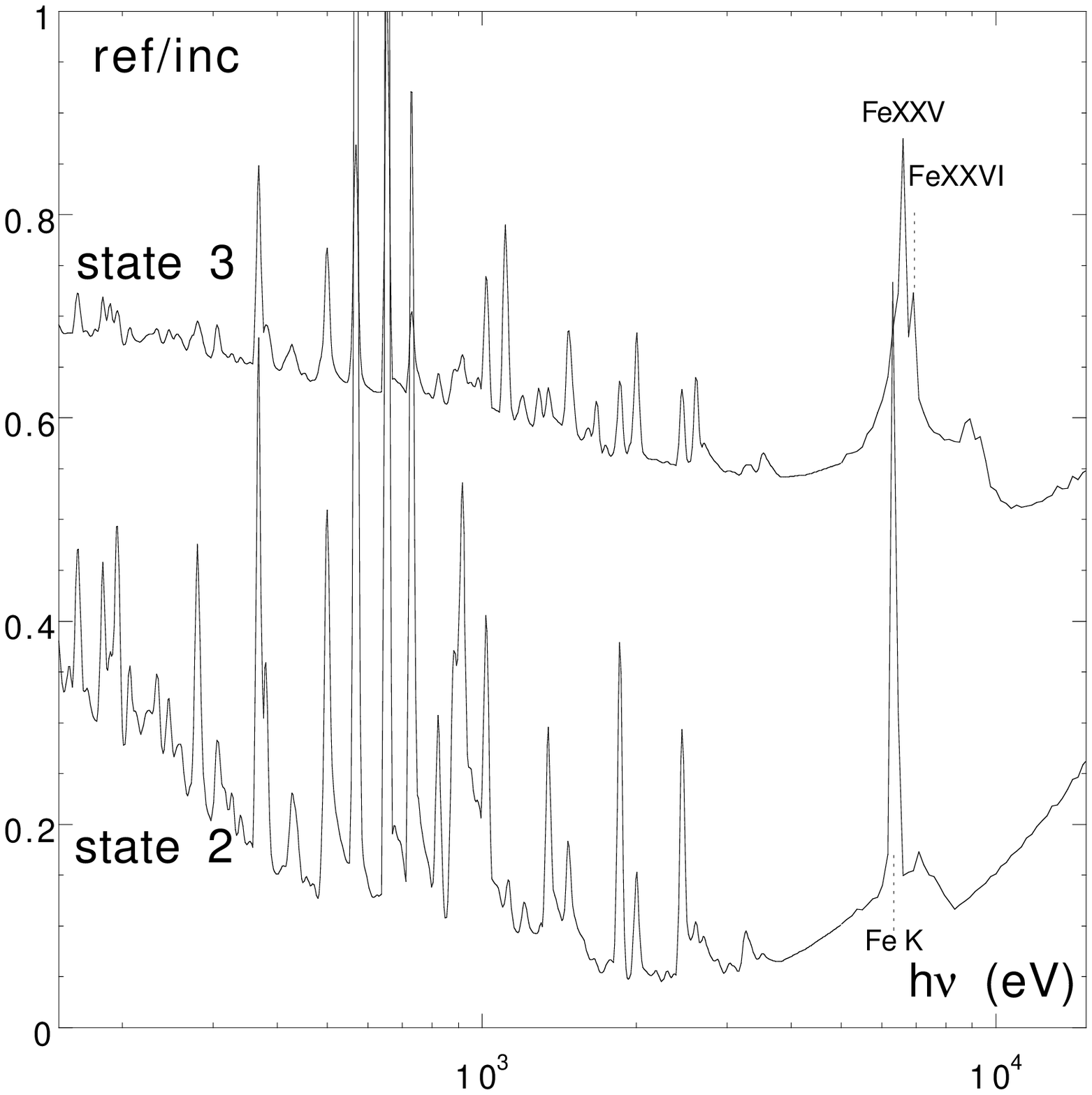,width=9cm}
\psfig{figure=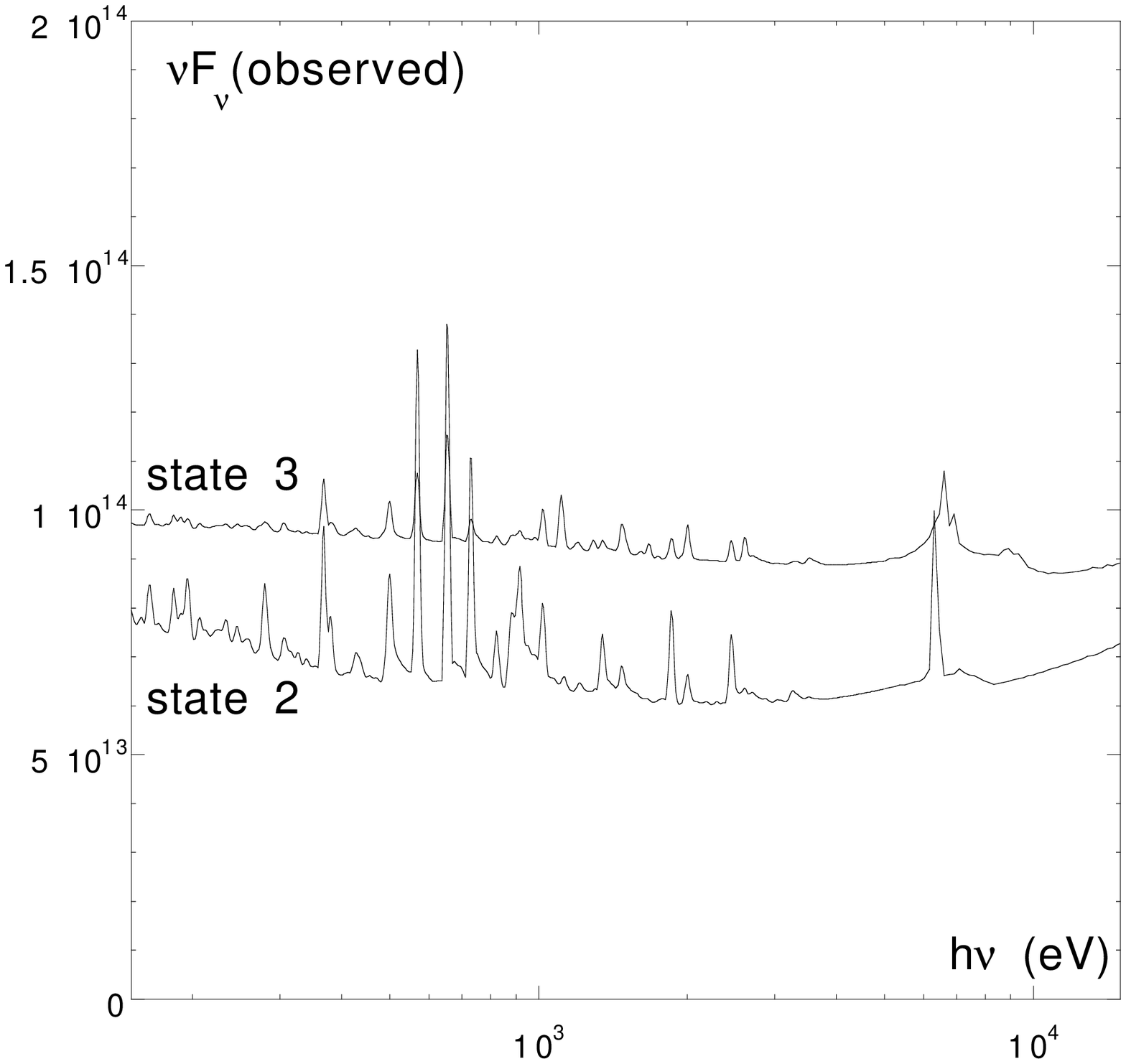,width=9cm}
\psfig{figure=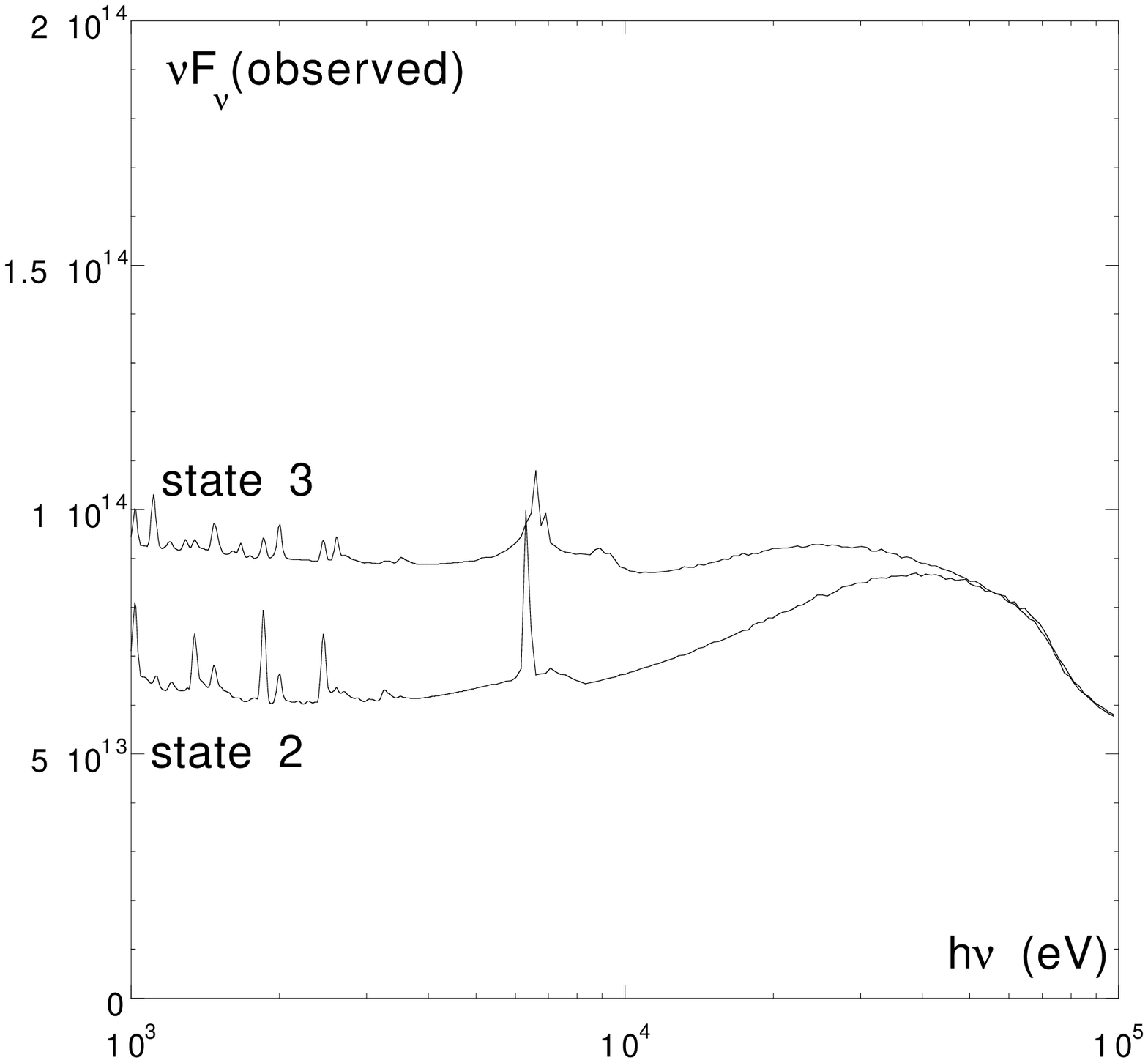,width=9cm}
\caption{Computed spectrum for model P1 (moderate flare), 
after onset of 
 the flare
(state 2) and 
after pressure equilibrium is reached (state 3).
The top figure gives the ratio of the reflected to the incident 
spectrum, and the two bottom figures give the sum of the two components 
(i.e. the observed spectrum) in two different spectral ranges. 
 The spectra are displayed with a spectral resolution of 44. }
\label{fig-model2-spe}
\end{center}
\end{figure}

\begin{figure}
\begin{center}
\psfig{figure=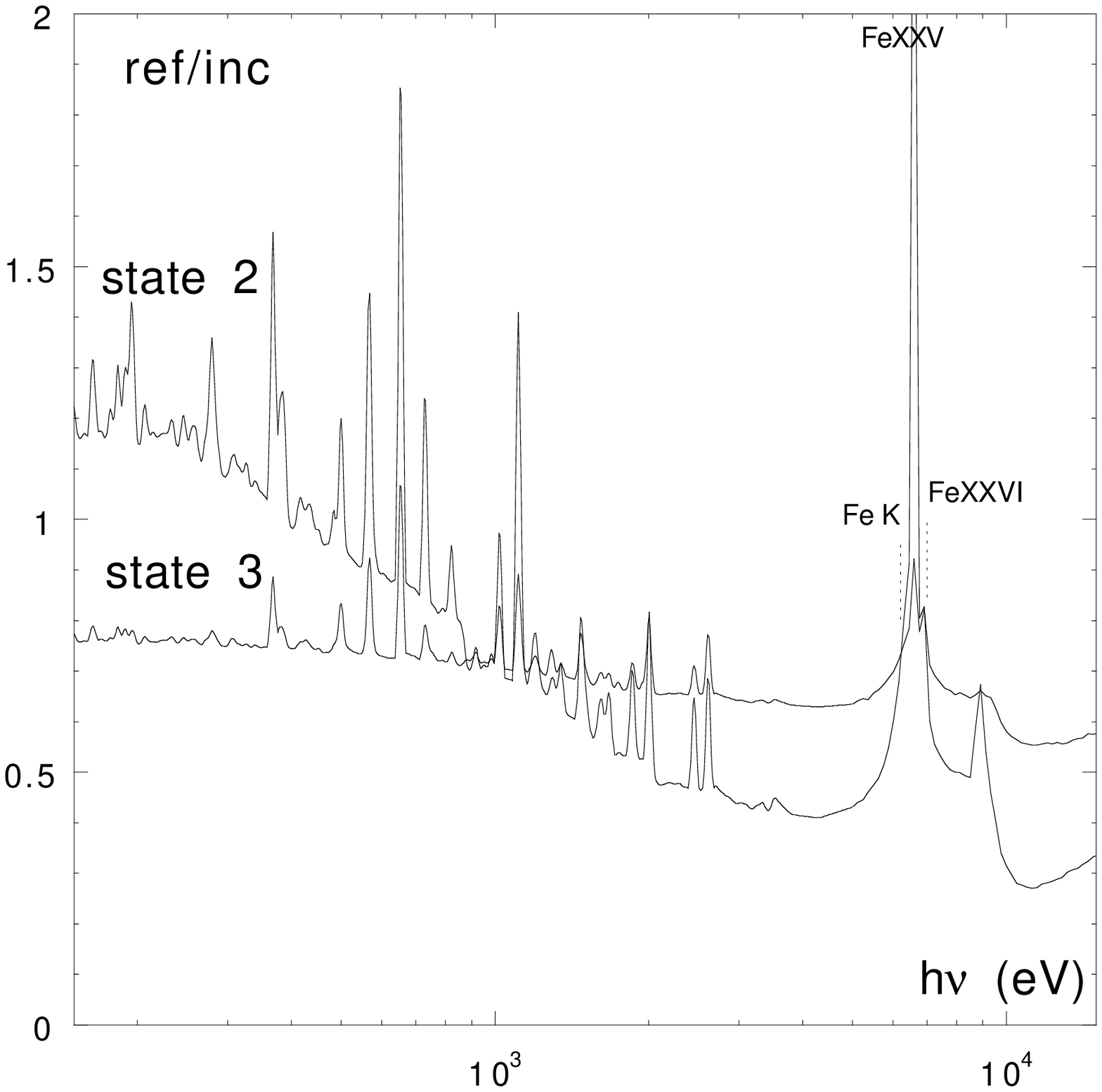,width=9cm}
\psfig{figure=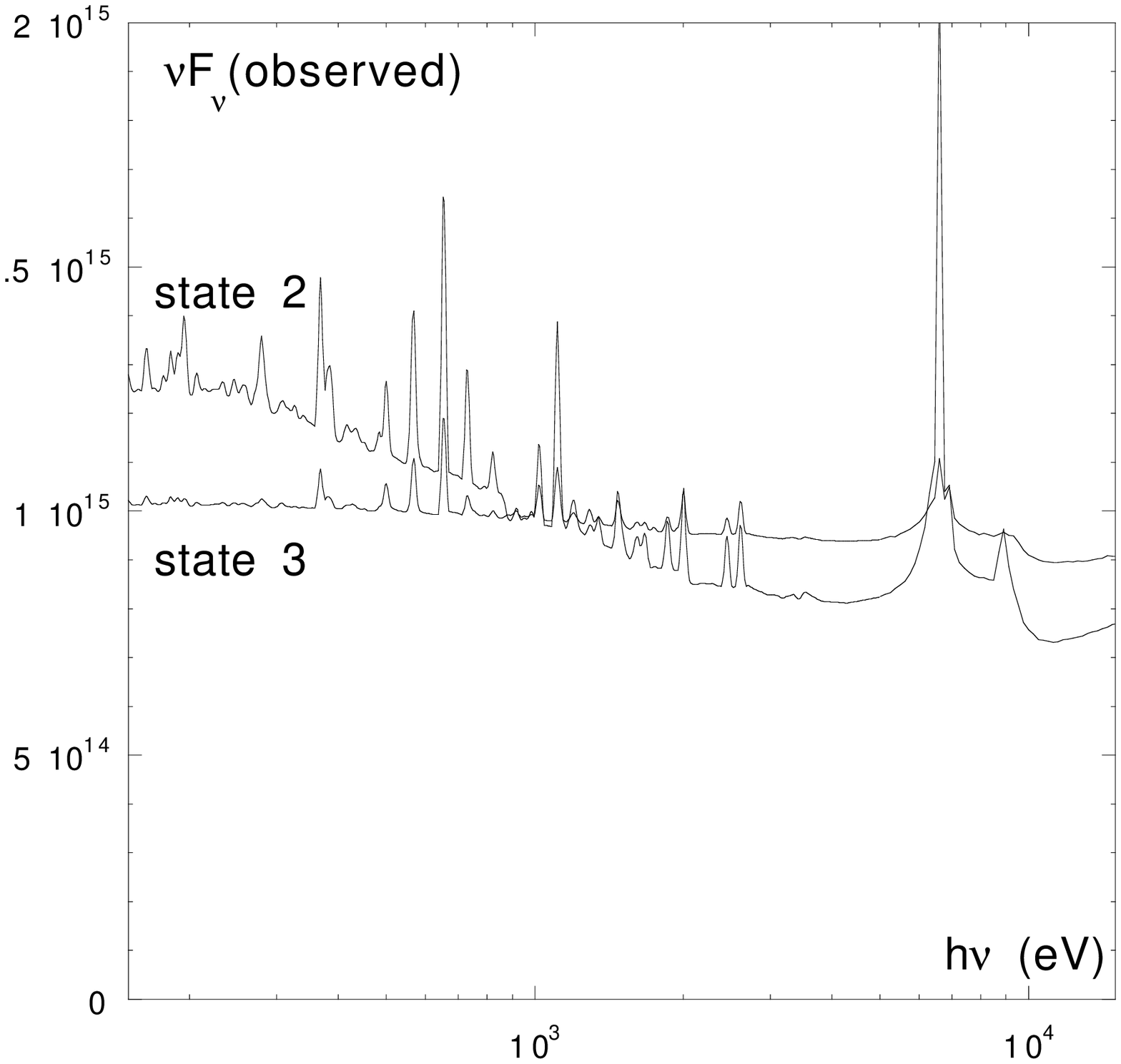,width=9cm}
\psfig{figure=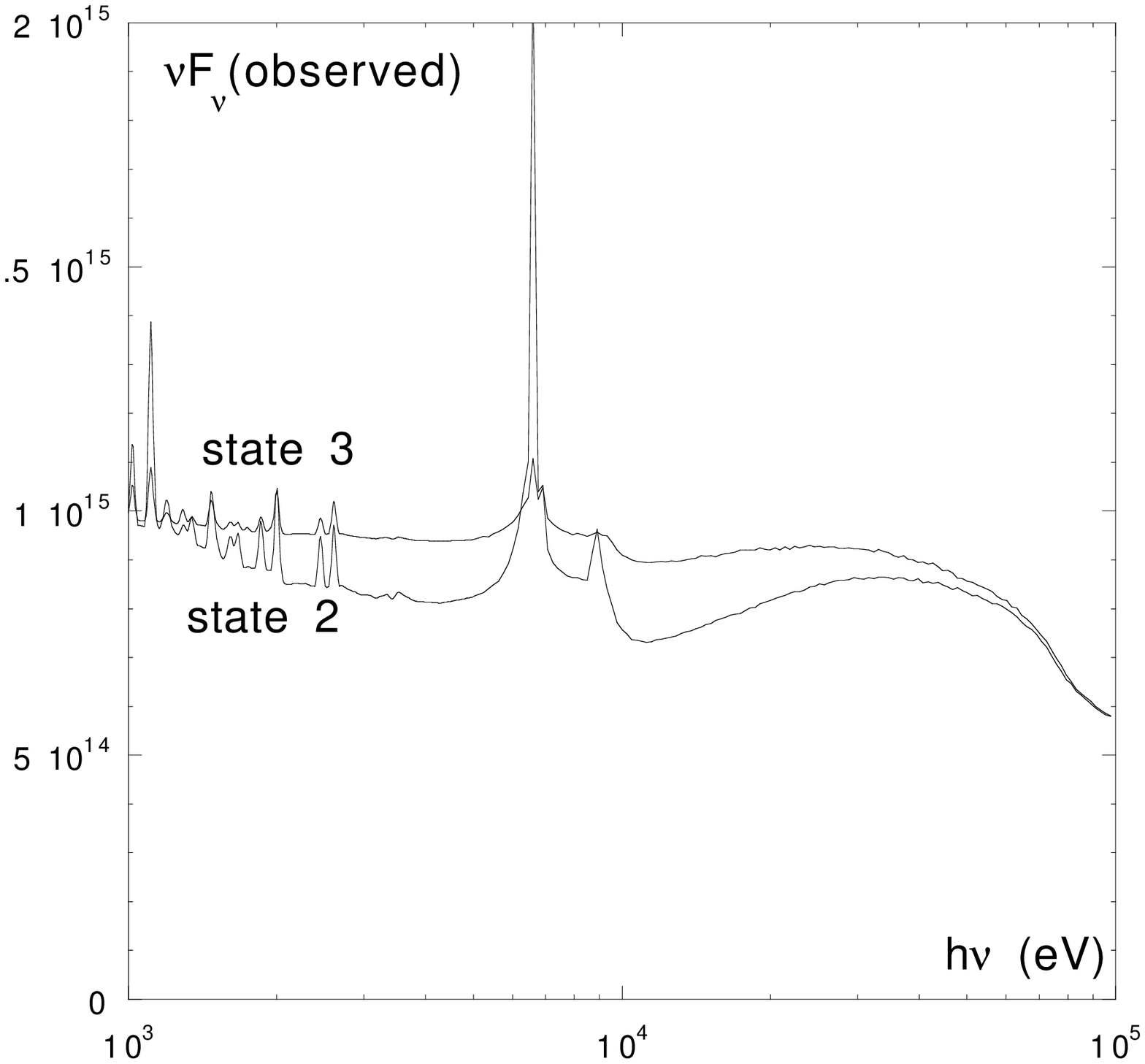,width=9cm}
\caption{Same as \ref{fig-model2-spe} for model P2 (strong flare).}
\label{fig-model3-spe}
\end{center}
\end{figure}

        We have performed two other tests, replacing the 
achievement of 
hydrostatic equilibrium by the achievement of {\it a constant total 
pressure}. 
 First it is not obvious that the only 
explanation for the properties of AGN is the ``standard irradiated 
disc" model in hydrostatic equilibrium, and other models have been proposed: very dense cloudlets in a spherical optically thin 
accretion flow (Rees 1987, Kuncic et al. 1997), or quasi-spherical 
distribution of more dilute clouds (Collin-Souffrin et al. 1996). 
Second, the disc model is strongly parameter-dependent. Its structure depends on 
the value of the viscosity parameter $\alpha$, 
on the 
vertical 
distribution of energy deposition, and even on the assumption made for the stress 
tensor (given by gas-pressure or by total pressure, or by a 
combination of both). None of these issues have yet been definitively 
established.  Assuming that the illuminated medium tends to settle at 
a constant pressure  avoids the
model-dependency of the results, and it leads to a significant
computational
simplification.  
Moreover this condition is not very different from hydrostatic 
equilibrium.  The geometrical thickness of the hot skin is at 
most equal to a few disc scale heights (cf. the appendix of R\'o\.za\'nska 
et al. 2002).
Thus the gravity does not vary much between the base of the 
atmosphere and the height corresponding to the bulk of the 
reflection (i.e. between $\tau_{es}$ = 0.01 and 1). 

Note also that there is an argument favoring a non-zero gas pressure at the surface 
of the non-perturbed accretion disc. 
The magnetic loops supposed to give rise to the 
flare are most likely located in a hot corona. The presence of this 
corona, 
even very dilute and contributing negligibly to the X-ray emission,
is felt by the disc as it exerts a gas pressure on the surface. 
The 
boundary condition is changed, and in particular the surface
density is increased. For instance in  the disc/corona model
studied by  R\'o\.za\'nska et al. (2002), the density in the disc 
atmosphere is constant and 
of the 
order of
$10^{15}$ cm$^{-3}$ in the whole atmosphere of the disc
(defined by $\tau_{es}\le 1$). So the density chosen in these test 
models ($10^{14}$ cm$^{-3}$) could well correspond to an atmosphere 
pressurized by a weak corona.

We have therefore used two models: model P1 corresponds to  a modest
 flare, and 
model P2 to a  
strong flare. 
In both models we assume that the unperturbed medium
is a slab 
of constant density $n=10^{14}$ cm$^{-3}$ and optical thickness 
$\tau_{es}=25$, illuminated after $t=0$ on one side by a flux which 
remains constant
  until pressure 
equilibrium is reached. The incident flux is $F_X=10^{15}\ 
{\rm erg\ cm}^{-2}{\rm s}^{-1}$ in model P1, $10^{16}\ 
{\rm erg\ cm}^{-2}{\rm s}^{-1}$ in model P2.
 The spectral distribution is a power 
law $F_{\nu}\propto 
\nu^{-1}$ from 0.1 eV to 100 keV (this spectral shape is chosen to 
allow 
comparison with the set of models already obtained in another 
context, cf. Coup\'e et al. 2003).
The backside of the slab is illuminated  by a blackbody of 
30000K, to mimic viscous dissipation, as was done for instance by 
Ross \& Fabian (1993). It has
 no influence on the reflected X-ray spectrum, 
since the 
corresponding viscous flux is 
much smaller than $F_X$. The total pressure of the final state   
is constant and equal to the gas pressure at the surface after the 
onset of the flare (cf. 
Figs. \ref{fig-P2-P} and \ref{fig-pcst-T-n}). This was assumed in order 
to mimic the situation of an 
accretion disk, where there is a high gas pressure underlying the hot skin which
 inhibits its expansion towards the back (actually 
there is a small mismatch in model P1, which has no consequence for the 
emission spectrum).

Fig. \ref{fig-pcst-T-n} 
displays the
temperature and density profiles, at time $t_{grow}$ after the onset of the 
flare 
(state 2), and after a time $t_{dyn}$, when 
pressure equilibrium 
is achieved (state 3) (state 1 before the onset of the flare 
corresponds to a low temperature 
 medium with almost no emission).

It is interesting to note that in the final equilibrium (state 3)
the temperature profiles
are almost the same for both models, as well as the optical thickness of the hot layer. 
This is due to the fact that the ``ionization parameter" (radiative 
flux to gas density ratio at the surface) is similar in both models. 
However the density is about 
one order of magnitude larger  
for model P2 in the deep layers, where  gas pressure 
dominates radiative pressure, since radiative pressure is one order of 
magnitude larger 
in the surface layers. 
The density on the illuminated side is not well
determined,  as gas pressure is completely negligible there. 
On the contrary, in the initial state 2, the 
optical thickness of the hot skin is almost one order of magnitude 
larger in model P2 than in model P1, as model P2 corresponds to a larger ionization 
parameter than model P1.  

 Note also that the 
temperature profile is smoother
when pressure equilibrium is not achieved, as in the hydrostatic 
case.

Figs. \ref{fig-model2-spe} and \ref{fig-model3-spe}
display the spectra for both models. 

The spectrum of model P1 looks like that of model H1, 
as expected, at 
least for the reflected to incident ratio. The ``observed" spectrum 
is slightly different, owing to the different spectral index of the incident 
continuum (the photon index $\Gamma$ is equal to 1.9 for model H1 and 
to 2 for models P1 and P2). The spectrum 
in state 2 is characterized by an intense Fe K$\alpha$ line at 6.4 keV,
 and a curvature 
of the continuum. 
When pressure equilibrium is achieved, the reflected continuum is 
flatter,
more intense, and the lines correspond to more ionized species, as in 
state 3 of model H1.

Model P2 leads in state 2 to a very steep reflected spectrum
 in the soft X-ray range (i.e. the albedo 
is larger than unity below 1 keV and smaller above 1 keV), while the emission 
lines (for 
instance the iron lines) correspond to highly ionized species (FeXXV 
and FeXXVI). It is 
because there is a large intermediate zone contributing to the 
reflection spectrum where the temperature decreases smoothly from
 10$^7$ K to  10$^5$ K. 
The spectrum in state 2 is 
 also characterized by intense lines and edges in the soft X-ray 
 range, but the iron line is dominated by  FeXXV lines, and the 6.4 keV line is absent. 
   In state 3, model P2 leads to a spectrum 
 similar to model P1, owing to their similar temperature structure. 

So we see that  {\it the structure and the spectrum are not 
very 
dependent on the power of the flare at pressure equilibrium. On the 
contrary, 
after the onset of the flare, for a strong flare
the soft X-ray spectrum is steep and the - very intense - iron line is dominated by highly 
ionized species, 
while for a moderate flare the X-ray spectrum is relatively flat and the - weak - iron line is dominated 
by neutral K$\alpha$ lines at 6.4 keV}. For 
instance the equivalent width of the iron complex is equal, in state 2, to 440 eV 
for model P2, and only to 104 eV
for model P1, while the EWs are almost the same in state 3 (100 and 78 
eV respectively). 

It is interesting also to compare the spectra obtained in the hard 
X-ray and gamma-ray range. While
Comptonization of the iron line is important in states 2 and 3 for the strong 
flare (model P2), it is negligible for state 2 of the moderate flare 
(model P1). In particular the moderate and the strong flare lead to a 
very different 
local 
iron line profile before pressure equilibrium is achieved.

\begin{figure}
\begin{center}
\psfig{figure=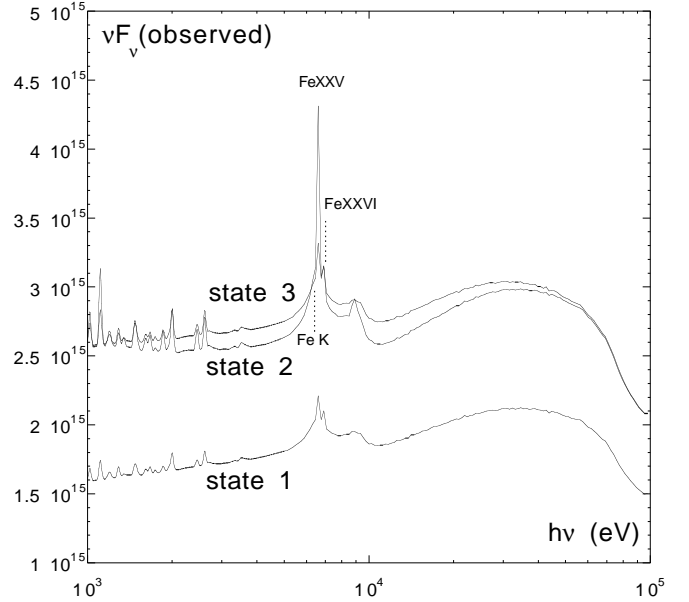,width=9cm}
\caption{Computed spectrum for states 1, 2 and 3 of combination 1
(cf. text for explanations). 
The figure gives the sum of the reflected + incident spectrum 
(i.e. the observed spectrum).  The spectra are displayed with a 
spectral resolution of 44. }
\label{fig-combin1-spe}
\end{center}
\end{figure}

\begin{figure}
\begin{center}
\psfig{figure=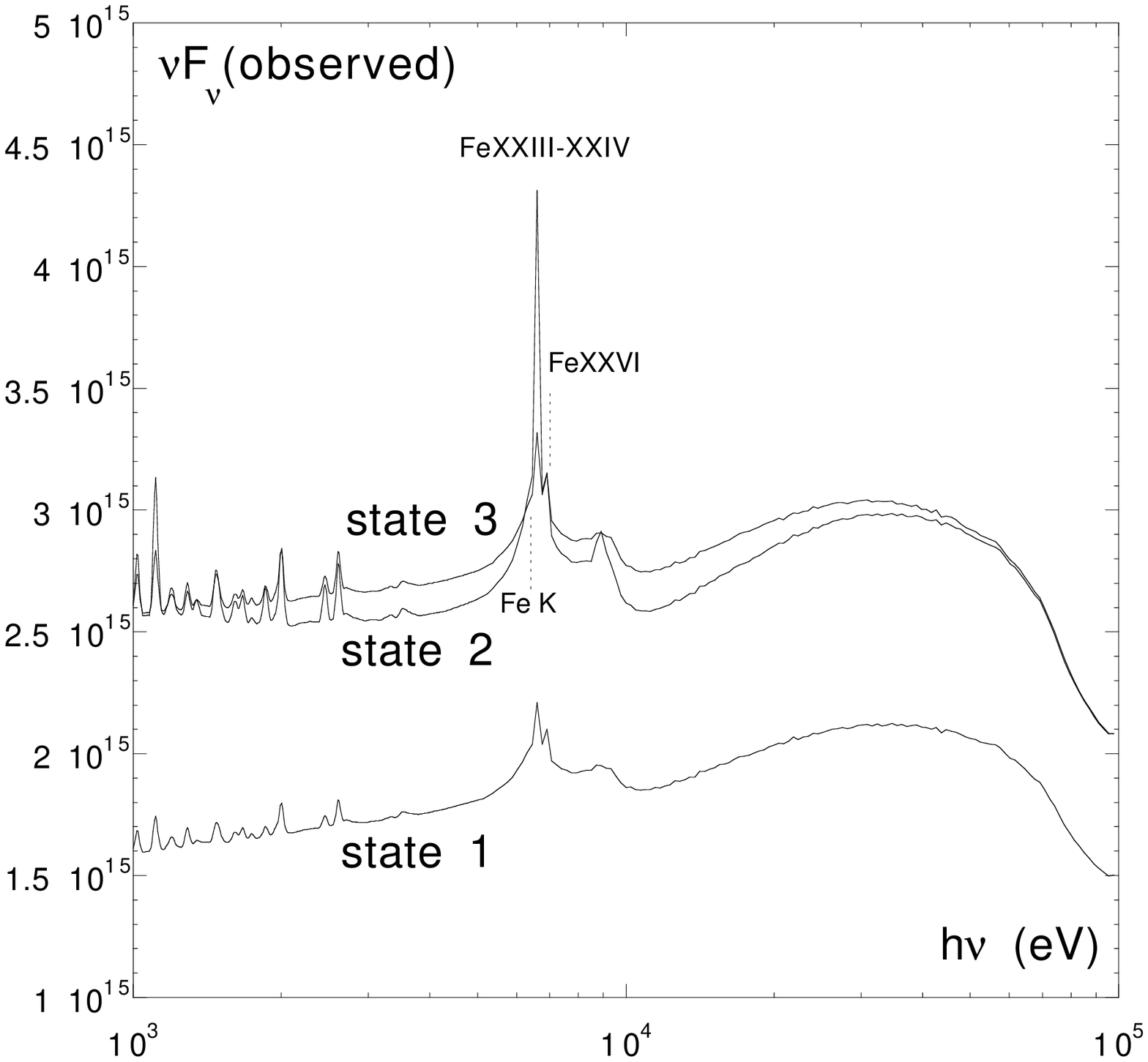,width=9cm}
\caption{Same as Fig. \ref{fig-combin1-spe} for combination 2. }
\label{fig-combin2-spe}
\end{center}
\end{figure}

\begin{figure}
\begin{center}
\psfig{figure=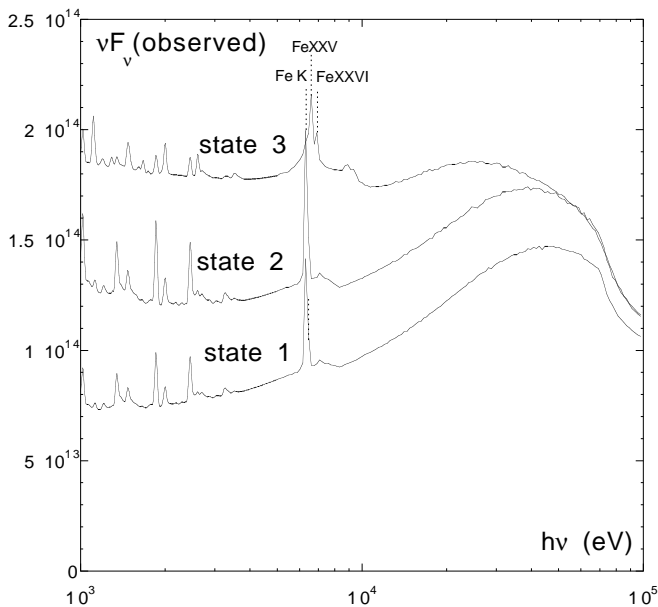,width=9cm}
\caption{Same as Fig. \ref{fig-combin1-spe} for combination 3. }
\label{fig-combin3-spe}
\end{center}
\end{figure}

\section{Discussion}

\subsection{Mixture of flares}

More than one flare can be present at the same time. However 
there are certainly not many of them, otherwise the variations would be 
completely erased
by the time delays and by the addition of many contributions. The 
observed spectrum should therefore be a mixture of a few flares in 
different states.  Actually, since 
the flares are independent, the observed spectrum is simply 
the addition of these different spectra weighted by the luminosity of 
each flare.

Fig. \ref{fig-combin1-spe} displays the observed spectrum 
for a combination of flares, assuming that all flares are similar 
to model H1 (combination 1). State 
1 corresponds to the addition of two 
flares having reached their hydrostatic 
equilibrium: i.e. state 1 of combination 1 = the addition of two states 3 of
 model H1.
 A third
 flare begins to shine. After a time $t_{grow}$, the ensemble reaches 
 state 2 of combination 1 that equals the addition of two states 3 of model H1, 
 and one state 2 of model H1. Finally, after a time $t_{dyn}$
the third flare also  reaches hydrostatic equilibrium; it is  state 3 of 
combination 1 = the addition of three states 3 of model H1. 

We see in Fig. \ref{fig-combin1-spe} that the increase of the flux due 
to the new flare
is linked to the appearence of lines in the 
soft X-ray range, and that of 
the Fe K$\alpha$ line at 6.4 keV, visible as a shoulder in the red 
wing of the line. These lines should slowly disappear as the disc
 reaches hydrostatic 
 equilibrium below the flare. 

Fig. \ref{fig-combin2-spe} displays the observed spectrum for another
combination of flares (combination 2).  State 1 of combination 2 
corresponds to the
addition of ten flares similar to model H1, having reached their
hydrostatic equilibrium (state 3 of model H1). A new flare begins to shine, 
but now it is a strong one,
 similar to
model P2: i.e. state 2 of combination 2 = the addition of ten states 3 
of model H1 and of one state 2 of model P2.  Again, the increase of the flux
 is linked with the apparition
of intense lines in the soft X-ray range, but contrary to
the previous case, the iron line at the beginning of the flare is 
already due to highly ionized species. The intensities of all these 
lines decrease when hydrostatic
equilibrium is reached. 

\medskip
 
Until now, we have only considered a simple change of flux during a
flare, not accompanied by a change of the spectral shape. However there
is growing observationnal evidence for a large number of Seyfert, that the
X-ray spectrum softens as the 2-10 keV flux increases (Perola et
al. 1986; Yaqoob et al. 1993; Lee et al. 1999; Lamer et al. 2000; Chiang
et al 2000; Petrucci et al. 2000; Vaughan \& Edelson 2001; Zdziarski \&
Grandi 2001; Georgantopoulos \& Papadakis 2001). We have mentioned that
Merloni \& Fabian (2001) interpret this effect as being due to the development
of an avalanche. 

We have therefore also considered  a third combination, which is the case of a flare
beginning with a relatively hard spectrum ($\Gamma$=1.9). When 
 the luminosity increases, the
incident continuum softens, reaching $\Gamma$=2.  The first flare has the
same parameters as the moderate flare of model P1, except that
the spectral index of the incident continuum is equal to 1.9. When 
 thermal equilibrium,  but not pressure equilibrium, is achieved, its 
 spectrum is
basically similar to state 2 of model P1, except that the slope of the
continuum is harder. It corresponds to state 1 of 
combination 3.  Then the luminosity doubles as the spectral index
increases and the flare becomes similar to model P1: it is state 2 of 
combination 3, equal to the addition of 
two 
 states
2 of model P1. State 3 of combination 3 corresponds to the achievement of 
pressure
equilibrium and therefore to the addition of two states 3 of model P1. 
Fig. \ref{fig-combin3-spe} displays the observed spectrum for combination 3.
 In the next section we will show the interest of this model, where
the increase in luminosity is accompanied by a softening of the 
spectrum.

\subsection{Comparison with the observations}

\subsubsection{Correlations between continuum parameters}

A correlation between the amplitude of the reflection component $R$ ($R$
is normalized so that $R=1$ in the case of an isotropic source above an
infinite reflecting plane) and the photon index $\Gamma$, has been found
by Zdziarski et al. (1999) in a large sample of Ginga
spectra of Seyfert galaxies. The measured $R$ tends to be larger in
softer sources.  This correlation is observed in different samples of sources as
well as in the time evolution of individual sources.  The $R$-$\Gamma$
correlation is also found in the {\it BeppoSAX} data (Matt 2001, and 
Perola et al. 2001), the
{\it BeppoSAX} sample showing however a significantly different shape,
with on average higher $R$ values at low $\Gamma$ and a flatter
correlation.  This correlation is still widely controversial, in
particular regarding the uncertainties on the measurement of $R$.
However no-one could prove that it is an artefact. The usual
interpretation of the correlation invokes the feedback from reprocessed
radiation emitted by the reflector itself. Very approximately, if 
such a correlation exists, one can represent it as $\delta R\propto 4 
\ \delta \Gamma$ 
for $\Gamma$ of the order of 2.

As mentioned previously, a correlation between $\Gamma$ and the X-ray
flux is also observed in Seyfert galaxies. We note however that
the fluxes used are generally restricted to a narrow energy band, usually
2-10 keV, and it is not clear if the total X-ray luminosity would follow
the same correlation (Petrucci et al. 2000). One observes an increase
of $\Gamma$ of 0.2 for an increase of the flux
by a factor 2 (cf. the fit of the correlation 
$\Gamma$-Flux of MCG -6-30-15 of Merloni \& Fabian 2001).

\begin{figure}
\begin{center}
\psfig{figure=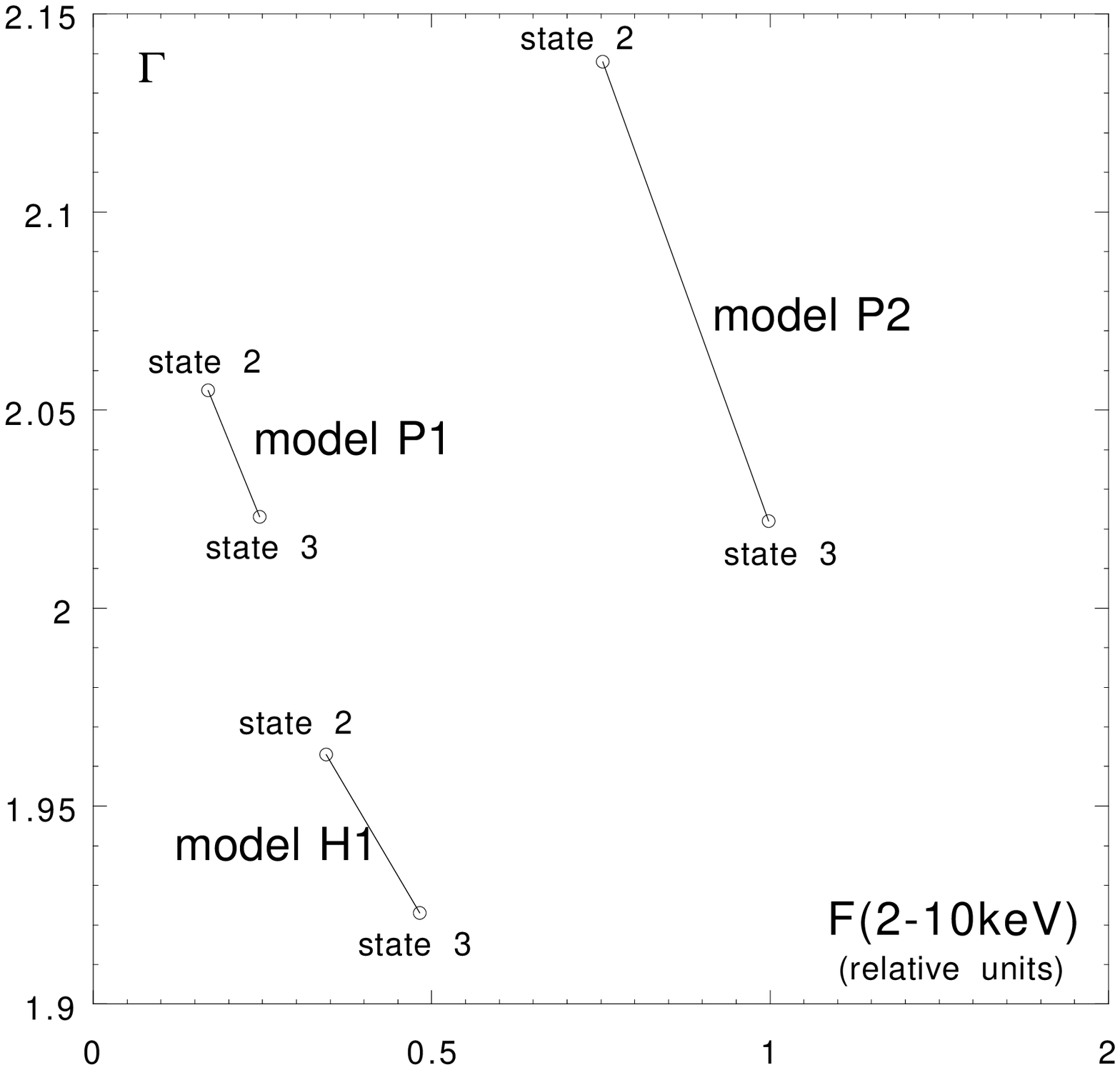,width=7cm}
\psfig{figure=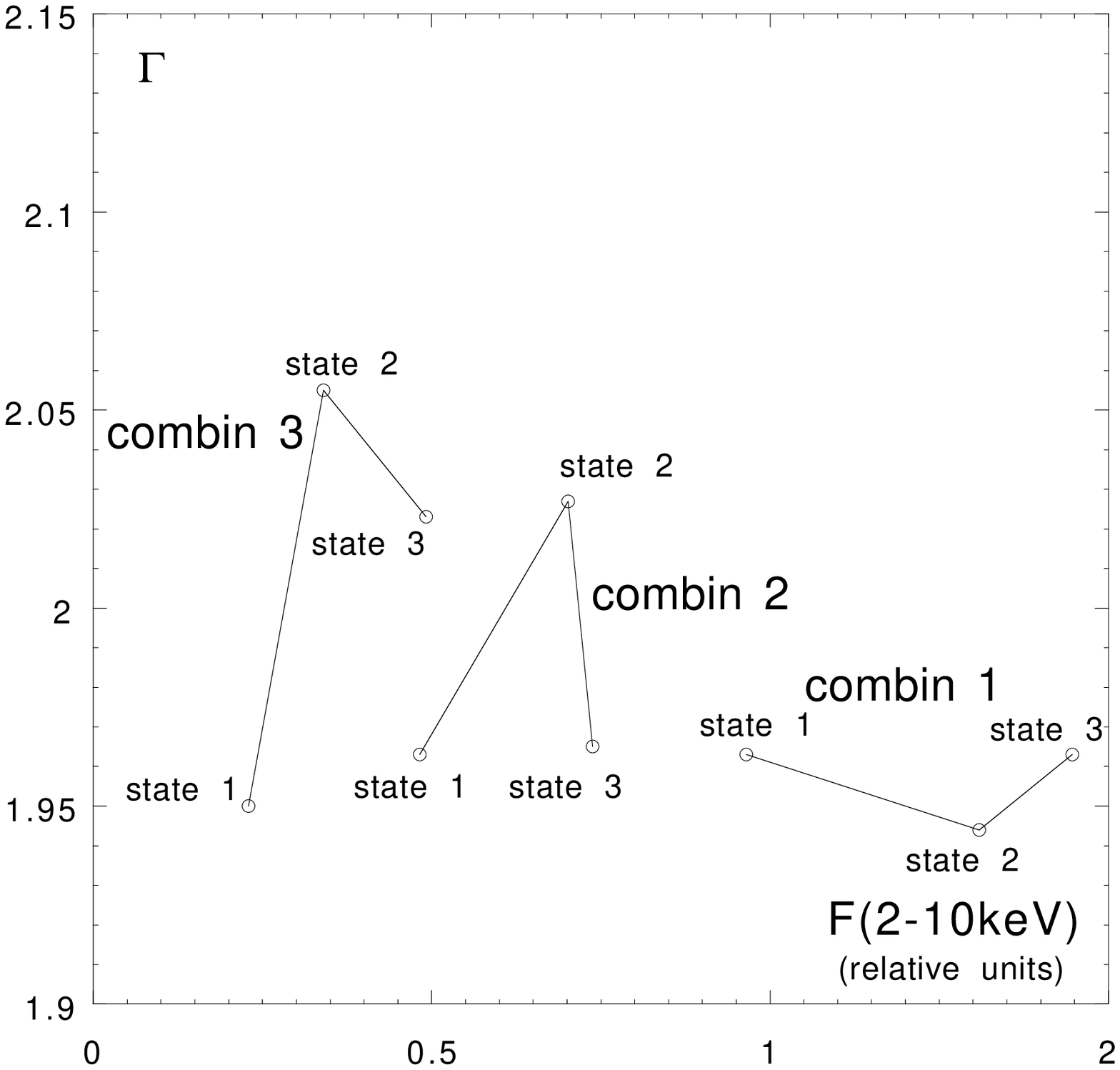,width=7cm}
\caption{Correlation between $\Gamma$ and $F$(2-10keV) expected during a flare, and deduced 
from the fits of the spectra. See text for explanations.}
\label{fig-fitsPOP-gamma-F}
\end{center}
\end{figure}

\begin{figure}
\begin{center}
\psfig{figure=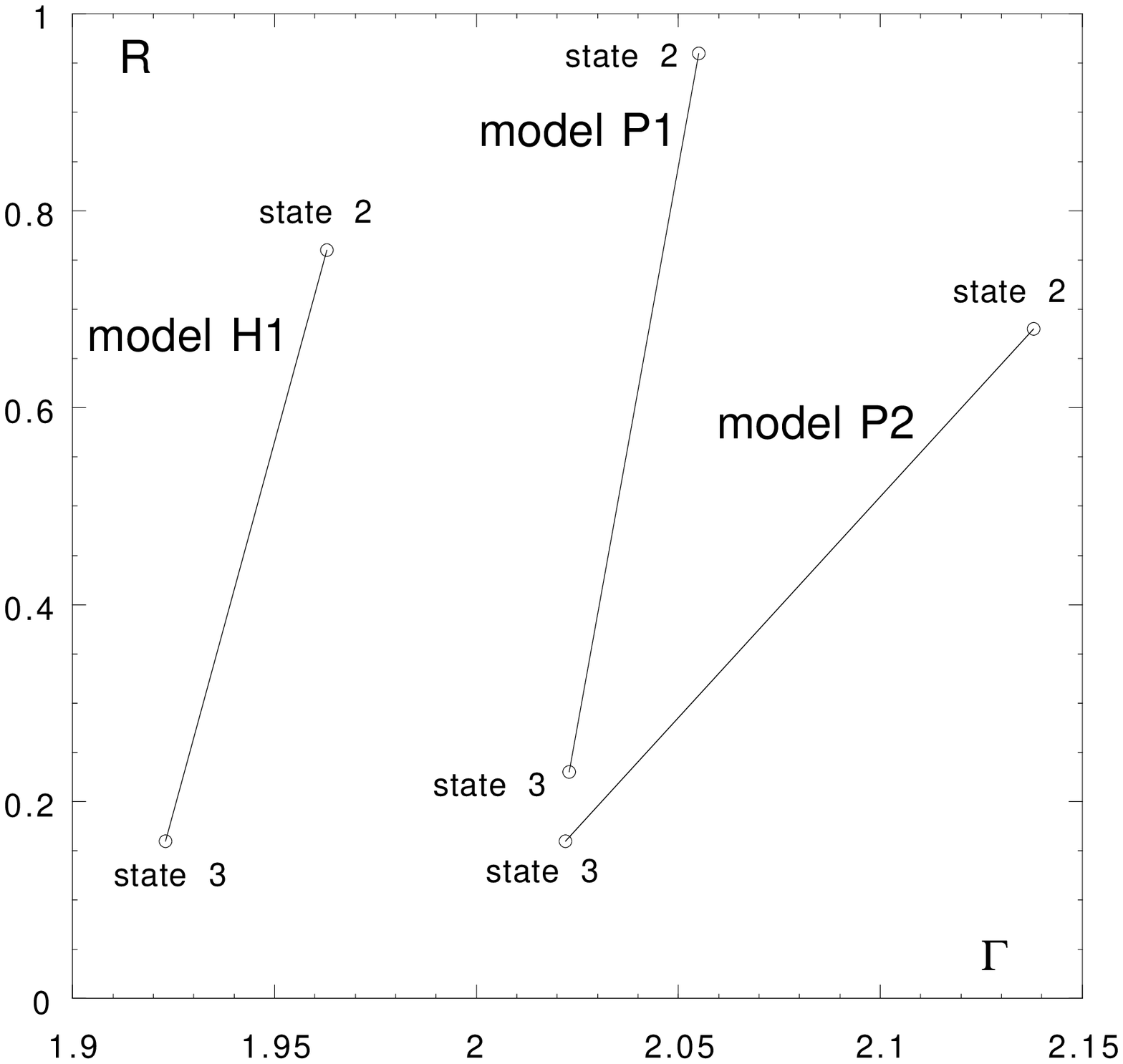,width=7cm}
\psfig{figure=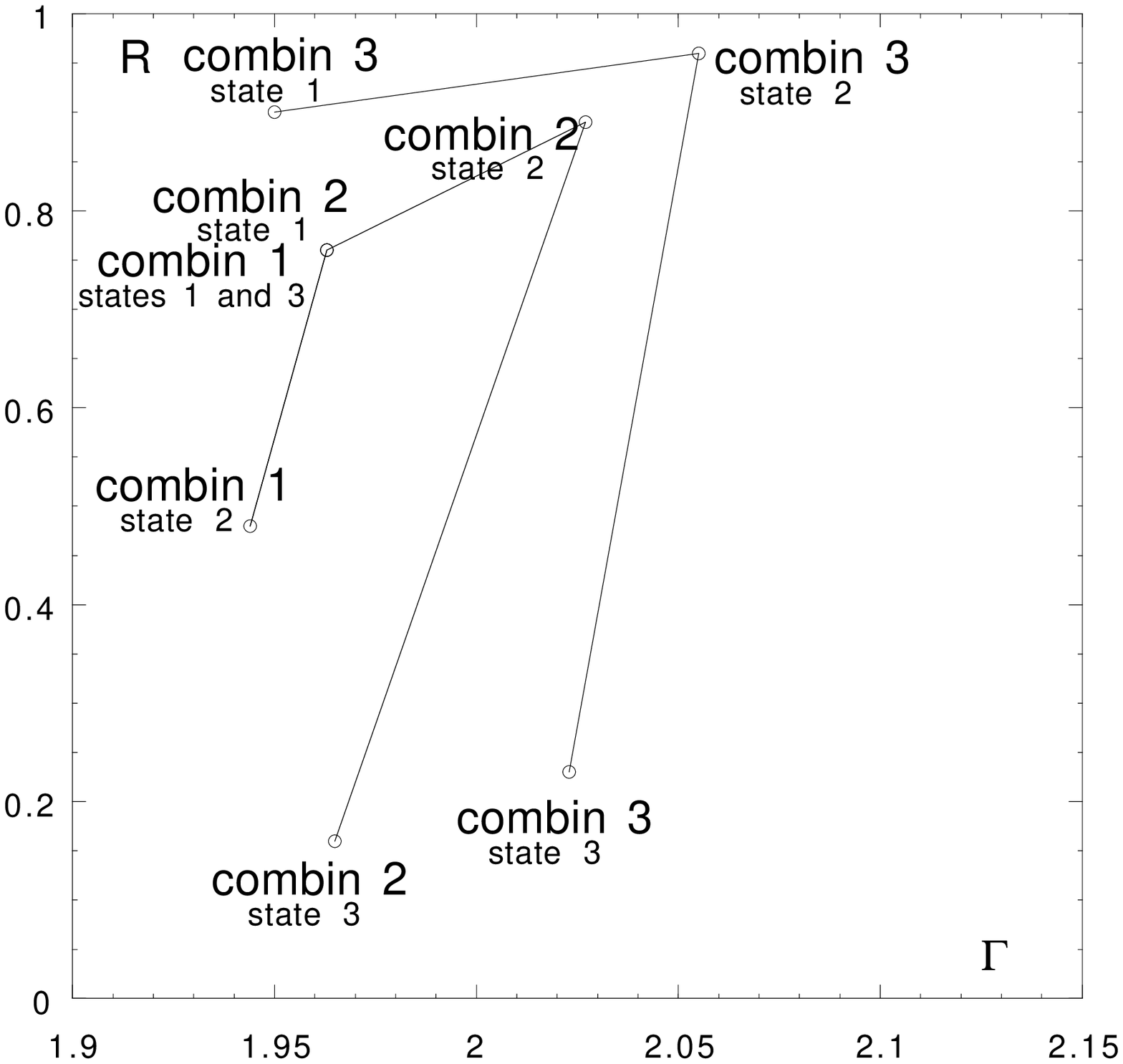,width=7cm}
\caption{Correlation between $R$ and $\Gamma$ expected during a flare, and deduced 
from the fits of the spectra. See text for explanations.}
\label{fig-fitsPOP-R-gamma}
\end{center}
\end{figure}

To see whether we can obtain these correlations with the flare model, we
have analyzed our different spectra with {\sc xspec}. For that purpose,
we needed to create, from our simulations, standard pha and unit diagonal
response matrix files using the task {\sc flx2xsp} of {\sc ftools}. We
assumed error bars of 5\% for each energy bin. We then fitted our spectra
with the {\sc pexrav} model of {\sc xspec} which has 3 parameters, the
photon index $\Gamma$, the reflection normalization $R$ and the high
energy cut-off $E_c$. Our spectra being limited to the 0.1-70 keV energy
band, we fixed $E_c$ at 400 keV for all fits (we have checked that, 
even
if other values of $E_c$ could result in small changes of our best fit
values and especially $R$, their relative trends between the different
states, which is what we are interested in, is insensitive to the value
of $E_c$). The fits have been made in two energy ranges, from
100 eV to 15 keV, and from 100 eV to 70 keV. 
The results for the two fits
are in good agreement.

Fig. \ref{fig-fitsPOP-gamma-F} shows the correlation between $\Gamma$ 
and $F$(2-10keV) computed with the different individual models (top figure)
 and the different 
combinations (bottom figure). It is clear that none of the individual models agrees
with the observations, i.e. they do not show an increase of the 
spectral index correlated with an increase of the flux. On the 
contrary, some combinations account well for the correlation: 
combination 1, between states 2 and 3, 
combination 2, between states 1 and 2, i.e. during the onset of a 
flare superposed on several old flares, and combination 3, also between states 1 and 2, i.e. during
 the 
onset of a moderate flare, 
 where it is assumed
that the slope of the incident X-ray spectrum itself increases during the
increase of luminosity.  A likely explanation for the lack of agreement between the
observations and the individual models could be that the flares do not last long
enough for the atmosphere to achieve hydrostatic equilibrium, and that
another flare begins before the atmosphere below the previous one has
reached pressure equilibrium.  In this case, the observed spectrum would
be dominated by several flares in non-equilibrium states, and would
display short-term variations (on $10^{3}-10^{4}$ s), similar to states 
1 and 2 of combinations
2 and 3. This result  
suggests that {\it the luminosity is always dominated by regions of 
the disc
having not reached their pressure equilibrium}.

Fig. \ref{fig-fitsPOP-R-gamma} shows the correlation between $R$ and $\Gamma$. 
A correlation between the two parameters is visible on all the 
individual models and the combinations. 
It is similar to that observed by
Zdziarski et al. (1999), even if our best fit values do
not completely agree with that reported by these authors. This could be
explained in part by the fact that our simulations suppose a simple power
law up to 100 keV for the continuum while we fit our spectra assuming
$E_c$=400 keV. It is not a critical point in our study however, since the
presence of a remote reflector (a torus) far from the central engine,
which has not been taken into account in our simulations, could easily
reproduce the $R$-$\Gamma$ correlation reported in the literature (Malzac
\& Petrucci 2002).

\subsubsection{Line spectrum}

 Nayakshin \& Kazanas (2001) stressed that the lines 
 emitted during a flare should be narrow, owing to the small dimension 
 of the active region, and they
should move in the frequency space according to the rotation of the 
disc. The shift in frequency would prevent one from 
distinguishing between a neutral and an ionized iron line if there is 
only one line. But if 
several ionization states are present, the frequency 
ratios would remain constant, and {\it the frequency shift would 
constitute a specific signature of a flare}.

The high sensitivity of the XMM-Newton satellite has recently permitted
the detection of narrow and rapidly variable iron lines.  Petrucci et
al. (2002) report a highly variable iron line in Mkn 841, unresolved with
the XMM-pn instrument, which disappears in about 15 hours. Turner et
al. (2002) claim, also from XMM-pn observations, the presence of
several variable narrow components within the profile of the iron
$K_{\alpha}$ line in NGC 3516. The presence of a varying narrow line
component in MCG -6-30-15 has also been detected by Lee et al. (2002) in
Chandra HETGS observations.  In the two latter cases, the line
variability appeared to be related to changes in the continuum flux. The
lines have been interpreted by the different authors as the narrow red or
blue wings of relativistically broadened profiles. It thus agrees with the
interpretation that magnetic flares, such as those studied here, may provide
the source of enhanced continuum emission by illuminating small areas of
the accretion flow. The flares have then to be produced near the central
black hole to explain the relativistic broadening.

In the case of Mkn 841, there were apparently no or only small changes of the X-ray
continuum when the iron line decreased or even disappeared. An 
interpretation which could be proposed is that the line observed 
during the first observation was due to a very strong flare still out of pressure 
equilibrium. When 
pressure equilibrium was achieved (which could have happened
during the time between the two observations), the line became 
very weak, while the continuum did not increase more. This would be 
similar to our model P2, except that the flare would have to be even 
stronger. New observations are clearly needed to confirm the line
variability in this source. We note however that it may also be produced
by different processes (like the presence of a concave disc as proposed by
Blackman 1999) which do not necessarily require variations of the
continuum.

Concerning the soft X-ray spectrum, it is difficult presently to 
perform a detailed comparison with the existing spectra, such as that 
of MCG -6-30-15. This would require the presence of a warm absorber in 
addition to the disc itself,
and to take into account Comptonization in the lines, which would smear the 
profiles, especially when pressure equilibrium is achieved.
 We intend to perform such computations in the near future.

\section{Conclusion}

We have made a preliminary study of the consequences of 
the flare model on the X-ray spectrum and on its time variations. 

We 
have in particular stressed the importance of the transient state during 
which the flare reached its maximum 
intensity and the irradiated atmosphere below has adjusted its thermal and 
ionization equilibrium, but is not in pressure equilibrium.  Such states 
are characterized by intense soft X-ray lines, and in the case of 
strong flares, by a very intense FeXXV line at 6.7 keV with a neutral component 
at 6.4 keV. 

We have also reached the conclusion that the observed correlation between 
$F$(2-10keV) and $\Gamma$ can be explained without invoking a 
feed-back of the reprocessed variation on the primary one. 
However an individual flare alone is not able to provide this 
correlation, as the simple variation of 
albedo occuring during the transition towards pressure equilibrium
leads to a 
correlation 
inverse to the observed one, and in particular to a steep soft X-ray spectrum.
 The onset of a strong 
flare superposed on several preexisting flares is required.  
This strongly suggests that the luminosity is dominated 
by transient states having not reached pressure equilibrium. 
On the other hand the correlation between $\Gamma$ and $R$, if 
real, is easily accounted for by all models. 

In the absence of pressure equilibrium,
 such atmospheres are not subject to the usual thermal instability, 
that leads to the disappearance of the intermediate temperature 
 and intermediate states of ionization. As a consequence the spectrum 
 displays both intense neutral and ionized iron lines.  
 Quite strangely, it would mean that the first models proposed for 
 irradiated discs, and
 assuming a
constant density illuminated slabs (Ross \& Fabian 1993, and 
susbsequent papers) are perhaps more relevant than the more sophisticated ones 
with atmospheres in hydrostatic equilibrium!

In this study, we have not taken into account any relativistic 
effects, and we have considered only {\it local} profiles. This is 
justified by the fact that the emission is provided by small localized 
regions on the disc. Thus the line profiles are not broadened by 
relativistic and gravitational effects, but only by Comptonization 
(which actually might be important, see the figures of the paper). On 
the other hand, the lines may be boosted, and
 red or blueshifted, and they could move according to the orbital motion 
 of the disc, if the emission region is located close to the black 
 hole.
 
 \medskip 

Several aspects should therefore differentiate the flare from the lamppost 
model:
\begin{itemize}

\item a shorter variation time scale, of $10^{3}-10^{4}$ 
s,

\item larger spectral variations during the increase of the  flux, including 
the appearance of 
 lines and edges
in the soft X-ray range and of very intense Fe K$\alpha$ lines at
 6.4 and 6.6 keV,
 
\item narrow and moving lines. 
 
\end{itemize}

We are not able to reach definitive conclusions 
about the reliability of either the lamppost or the flare model in the 
context of present variation studies, though a few observations tend to support 
 the flare model. More 
observations are required to understand the relationship between the 
flux, the
iron line, the soft and hard X-ray continua.
To make some progress, it is also necessary to compare a larger 
set of models to the observations.  The aim of
 this paper was simply to draw  attention to the differences between 
 the lamppost and the flare model, in particular to the
 fact that the disc underlying a flare is ``cold"
before the flare settles, and more generally to the fact that the 
luminosity is certainly dominated by transient states out of pressure 
equilibrium. We have given here only a few  typical 
 examples, postponing a more 
in-depth study to a future paper.

\begin{acknowledgements}

We are grateful to Martine Mouchet for a careful reading of the manuscript, 
which led to substantial improvements.

Part of this work was supported by grant 2P03D01816 of the Polish
State Committee for Scientific Research and by Jumelage/CNRS No.\ 16 
``Astronomie France/Pologne''.

\end{acknowledgements}

\end{document}